\documentclass[%
 reprint,
 amsmath,amssymb,
 aps,
]{revtex4-2}
\usepackage{physics}
\usepackage{float}
\usepackage{graphicx}% Include figure files
\usepackage{dcolumn}% Align table columns on decimal point
\usepackage{bm}% bold math
\usepackage{amsmath}
\usepackage{witharrows}
\usepackage{hyperref}

\def\taurel{{\tau_{\text{relax}}}}
\usepackage[normalem]{ulem}

\begin{document}

%Shortcut to Thermal Equilibrium: 
\title{Analytical Results and Universal Behavior in 
Fast Thermal Equilibration Protocols }% Force line breaks with \\
%Unveiling Analytical Solutions: Two-Step Protocol for Shortcut to Equilibrium and Universal Behavior in Thermodynamic Systems
\author{Diego Rengifo}
\email{d.rengifo@uniandes.edu.co}
 %\altaffiliation[Also at ]{Physics Department, Universidad de los Andes}%Lines break automatically or can be forced with \\
\author{Gabriel T\'ellez}%
 \email{gtellez@uniandes.edu.co}
\affiliation{%
 Physics Department, Universidad de los Andes, Bogot\'a, Colombia.\\
}%


\begin{abstract}

When a system deviates from equilibrium, it is possible to manipulate and control it to drive it towards equilibrium within finite time $t_f$, even by reducing its natural relaxation timescale $\taurel$. Although numerous theoretical and experimental studies have explored these shortcut protocols, few have yielded analytical results for the probability distribution of the work, heat, and produced entropy. In this study, we propose a two-step protocol that captures the essential characteristics of more general protocols and provides an analytical solution for the relevant thermodynamic probability distributions. Additionally, we present evidence that for a very short protocol duration $t_f\ll \taurel$, all protocols exhibit universal behavior for the ratio of probability distribution functions of positive and negative work, heat, and the produced entropy. 
\end{abstract}

\maketitle

\section{Introduction}

Almost all the thermodynamic systems in nature are out of equilibrium. Equilibrium states are, therefore, not common but desirable. When a system is out of equilibrium and left without any external intervention, it takes (technically) an infinite time to reach an equilibrium state. The speed at which the system approaches the final equilibrium state is characterized by timescale $\taurel$. This time is an intrinsic characteristic of any physical system and depends on various factors such as the underlying interactions, which are encoded in the friction, transport coefficients, external parameters (if any), and temperature~\cite{kubo2}.

Technology is advancing rapidly, and one of its trends is the creation and control of smaller devices. This miniaturization of devices has led to increasing interest in the development of engineered techniques that can shorten the natural timescale for relaxation between equilibrium states. These procedures are designed to connect the equilibrium states through a protocol that is considerably shorter than the natural equilibration time. Such techniques were inspired by the so-called shortcut to adiabaticity~\cite{UNANYAN97,chenSTA10, odelin1}. Since then, the term Engineered Swift Equilibration (ESE) has been coined to describe these protocols~\cite{trizac1}. These protocols are also known as the ``shortcut to isothermality"~\cite{geng17} or ``swift state-to-state transformations"~\cite{guery-odelinDrivingRapidlyRemaining2023}.

%(ESE REFERS TO PROTOCOL OR PROCESS...)
Several ESE protocols have been established, including those for frictionless atom cooling in harmonic traps~\cite{odelin2, odelin3} and nanosystems as micromechanical oscillators in contact with a thermostat, both in overdamped and underdamped regimes~\cite{trizac1, trizac2}. These procedures enable the creation of a nonequilibrium state that can be controlled and manipulated, thereby allowing the exploration of novel physical phenomena~\cite{guery-odelinDrivingRapidlyRemaining2023}.

Overall, the ESE protocols has shown great promise in the field of nonequilibrium statistical mechanics and has opened new avenues for research in the development of novel techniques for controlling and manipulating the dynamics of physical systems, especially nanodevices. Due to the increased miniaturization of chips, robots, and devices, understanding controlled dynamics has become mandatory in modern technology. To advance our understanding of how protocols accelerate system equilibration it is crucial to comprehend the stochastic thermodynamics of small systems. For this reason, it is vital to calculate the probability distribution functions (PDF) of work, heat and produced entropy for these distinct processes.

In this study, we consider a Brownian particle, the system, immersed in a viscous medium, and the environment at temperature $T$, whose particles are smaller than the Brownian particle. This justifies the validity of the commonly known overdamped regime, which is particularly applicable to colloidal particles. The particle is subjected to a time-dependent and externally controlled harmonic-type potential, which allows precise manipulation and control. An experimental realization of this system was made with colloidal particles trapped by laser beams~\cite{wang02,BlicklenonHarmonic06,trizac1} and has been used to build microscopic heat engines~\cite{BlickleBechinger12, martinezBrownianCarnotEngine2016}.

When the stiffness of the harmonic-type potential is controlled through ESE protocols, the probability distributions of the relevant thermodynamic quantities, such as the work and heat, cannot generally be calculated analytically. Therefore, their theoretical study is limited to numerical simulations. In this study, we introduced a two-step protocol as a novel approach to derive exact analytical results. Remarkably, the analytical solution of the two-step protocol captures the essential characteristics of more general protocols, exhibiting a certain level of universality in the behavior of the work, heat, and entropy probability distributions.

%In this article,  we will focus on a Brownian particle trapped in harmonic potential with time-dependent stiffness controlled externally. We will assume that the overdamped regime is valid, this is justified for colloidal particles

The remainder of this paper is structured as follows. In Sec.~\ref{sec:ESE_protocol}, inspired by the ideas presented in Ref.~\cite{trizac1}, we develop a protocol that establishes a connection between two equilibrium states, theoretically reducing the equilibration time. In Sec.\ref{sec:TSP}, we propose a simplified toy model called the Two-Step Protocol (TSP), which allows for an analytical solution of the probability distribution functions of relevant thermodynamic quantities.  In sections~\ref{sub_sec:work pdf}-~\ref{sub_sec:entropy}, we calculate the work, heat, and produced entropy probability distributions for the TSP. We also check the validity of the Jarzynski equality~\cite{Jar97, jarzynski_equilibrium_1997}, Crooks relation~\cite{Crooks1998, crooks_entropy_1999}, and entropy fluctuation theorem~\cite{Evans1} for this specific protocol. 

Finally, a comprehensive comparison with more general protocols is presented and reveals similar features in the ratio $P_A(A)/P_A(-A)$ of the probability distributions, where $A$ is the work, heat, or produced entropy. These relations differ from the usual fluctuation theorems~\cite{Evans1, Evans2, gallavottiDynamicalEnsemblesStationary1995, Jar97, jarzynski_equilibrium_1997, Crooks1998, crooks_entropy_1999}, as both $P_A(A)$ and $P_A(-A)$ refer to the probability distribution of the forward process.

%Here, we do not consider a backward process with the reversed time direction.
%A similar protocol have been  reported in \cite{trizac1} with its respective experimental realization. We follow some ideas exposed 

%The principal aim of this paper is to calculate the probability density function (PDF) for heat and work. However, obtaining the analytical expression for these functions is impossible for the protocol previous posted, as well as for the majority of other protocols \cite{speck}. As a result, we introduce a simplified model that incorporates the essential features of the broader problem and enables us to derive analytical solutions. The analytical solutions obtained from this model are particularly advantageous, as they can provide insights that extend beyond what numerical methods can achieve.

\section{ESE Protocol}\label{sec:ESE_protocol}
%\dr{ESE IS A PROCESS OR PROTOCOL AGAIN}
In this section, we review the concepts underlying ESE protocols~\cite{trizac1, guery-odelinDrivingRapidlyRemaining2023}. The central idea behind the ESE protocols is to construct a customized time-dependent protocol $\lambda(t)$ for the externally controlled parameter. This protocol is specifically designed to guide the system from the initial equilibrium state characterized by $\lambda_i$ to the desired final equilibrium state characterized by $\lambda_f$ within a finite time interval $t_f$ which is shorter than the system's equilibration time $\taurel$.
Let us consider a Brownian particle in a thermal environment at temperature $T$, and trapped in a harmonic potential with time-dependent stiffness, given by
\begin{equation}
U(x,t)=\frac{1}{2}k(t)x^2.
\end{equation}
Here, $x$ is the position of the particle, and protocol $\lambda(t)$ is essentially characterized by the stiffness $k(t)$.
Initially, the particle is in thermal equilibrium at temperature $T$ with the stiffness $k_i$. As the stiffness varies in time, the system evolves out of equilibrium. The goal is to design a protocol for the stiffness $k(t)$, such that the system 
reaches a final thermal equilibrium state with stiffness $k_f$ at time $t_f$. However, there are infinite possible protocols that can achieve this transformation; therefore, constraints can be imposed to obtain the desired solution. Since the goal is to reduce the equilibration time through external control, the problem falls within the realm of the optimization theory~\cite{trizac3, Bechhoefer_control21}.

A solution to this problem was presented in Ref.~\cite{trizac1}, where both experimental and theoretical results were obtained in the overdamped limit, that is, the acceleration term in the Langevin equation was neglected. Hereafter, we assume that the overdamped limit is valid. In this limit, the corresponding Langevin equation is given by
\begin{equation}\label{langevin}
\dot{x}=-\frac{k(t)}{\gamma}x+\sqrt{2\mathcal{D}}\xi(t),
\end{equation}
where $\xi(t)$ is Gaussian white noise with zero average and the autocorrelation function $\expval{\xi(t)\xi(t')}=\delta(t-t')$. Here, $\beta=1/(k_B T)$ and $\gamma$ are the inverse temperature and friction coefficient, respectively, and $\mathcal{D}$ is the diffusion constant $\mathcal{D}=k_B T/\gamma$.

%%% Discusión sobre unidades reducidas
%
%If the stiffness is fixed at a constant value $k_f$, the evolution is given by a Ornstein-Uhlenbeck process which has a natural relaxation time scale $\taurel=\gamma/k_f$. 
From the Langevin equation, we infer that the natural timescale is given by $\taurel=\gamma/k_f$. It is useful to use a set of dimensionless variables associated with this timescale and the final stiffness $k_f$ of the protocol as follows: $\tilde{t}=t/\taurel$, $\tilde{x}(\tilde{t})=x(t)/\sqrt{\mathcal{D}\taurel}$, and $\tilde{k}(\tilde{t})=k(t)/k_f$. The natural energy scale is $k_B T$ and is given by $\tilde{U} = U/(k_B T) = (1/2) \tilde{k}(\tilde{t}) (\tilde{x}(\tilde{t}))^2$. With this set of units, the reduced Langevin equation reads
\begin{equation}
\frac{d\tilde{x}}{d\tilde{t}} = -\tilde{k}(\tilde{t}) \tilde{x}(\tilde{t}) +\sqrt{2} \tilde{\xi}(\tilde{t}),
\end{equation}
where $\tilde{\xi}(\tilde{t})=\sqrt{\taurel} \xi(t)$ satisfies $\langle \tilde{\xi}(\tilde{t})\rangle = 0$ and $\langle \tilde{\xi}(\tilde{t})\tilde{\xi}(\tilde{t}')\rangle = \delta(\tilde{t}-\tilde{t}')$. Using these units, it is clear that there are only two parameters for our problem: the initial stiffness with respect to the final stiffness, $\tilde{k}_i=k_i/k_f$, and the target duration of the protocol compared to the relaxation timescale, $\tilde{t}_f=t_f/\taurel$. Henceforth, we use these dedimensionalized units and remove the tilde to lighten the notation.

%TAREA por hacer: Usar unidades adimensionales en todo el resto del texto y quitar las tildes para simplificar notación. Esencialmente hay que poner $\beta=1$, $\gamma=1$, $\taurel=1$. También se tiene $k_f=1$ pero se puede dejar escrito como $k_f$ por si uno decide usar como referencia otro valor de $k$.

%%%%

%\dr{and the goal of the ESE protocol is to reach a new equilibrium state at time $t_f$}

The system is in an equilibrium state at the beginning $t_i=0$. As a result, the position probability distributions can be described by Gaussian distributions, which are characterized by 
\begin{align}\label{PVI0}
    \expval{x(0)} &=0,\\
    \sigma_i^2=\expval{x^2(0)}&=\frac{1}{k_i},
    \label{PVIa}\\
    \sigma_f^2=\expval{x^2(t_f)}&=\frac{1}{k_f}=1.
    \label{PVIb}
\end{align}
As is well-known~\cite{risken}, the Langevin equation can be mapped to a Fokker-Planck equation 
\begin{equation}
\frac{\partial P}{\partial t}=\frac{\partial}{\partial x} \left[ k(t)xP\right]+\frac{\partial^2 P}{\partial x^2}
\end{equation}
for the probability density function $P(x,t)$ of position. To solve this equation, we performed a Fourier transform
\begin{equation}
    \hat{G}(p,t)=\int_{-\infty}^{\infty}P(x,t)e^{-ipx}\,dx
\end{equation}
leading to an equation of the form
\begin{equation}
\frac{\partial}{\partial t}(\ln \hat{G})=-k(t)p \frac{\partial}{\partial p}(\ln \hat{G})-p^2.
\end{equation}
The combination $\ln(\hat{G})$ is the cumulant generating function whose Taylor series is
\begin{equation}
    \ln{\hat{G}(p,t)}=\sum _{n=0}^{\infty}\chi_{n}(t)\frac{(-ip)^n}{n!},
\end{equation}
where $\chi_n(t)$ denotes the cumulants of $P(x,t)$. The average value and variance correspond to $n=1$ and $n=2$, respectively. The insertion of this expansion leads to the derivation of a set of ordinary differential equations governing the time evolution of all the cumulants for the position. Notably, the only non-trivial equation corresponds to $n=2$ ($\chi_2(t)=\langle x(t)^2 \rangle$)
\begin{equation}
    \dot{\chi}_2(t) + 2k(t)\chi_2(t)=2. \label{n2}
\end{equation}
The essence of ESE protocols is to propose a particular functional form for the variance $\chi_2$ such that it satisfies the initial Eq.~(\ref{PVIa}) and the final conditions in Eq.~(\ref{PVIb}): $\chi_2(0)=1/k_i$ and $\chi_2(t_f)=1/k_f=1$. Subsequently, the appropriate stiffness, $k(t)$ can be extracted from Eq.~(\ref{n2}). Thus, the process of determining the appropriate stiffness was performed in a reverse-engineered manner.

To ensure a smooth transition to equilibrium, two additional conditions can be optionally imposed on $\chi_2(t)$, given by
\begin{equation}\label{pviderivative}
\dot{\chi}_2(0)=\dot{\chi}_2(t_f)=0.
\end{equation}
We have four conditions that must be satisfied; hence, we need four parameters for adjustment. In addition, we aim to optimize the work done on the system during the process by adding a fifth parameter that allows us to tune it in such a way that the work is minimized. Based on this consideration, we propose a solution in the form of a fourth-degree polynomial for $\chi_2$

\begin{equation}\label{proposal}
\chi_2(t)=A_0+A_1t+A_2t^2+A_3t^3+A_4t^4.
\end{equation}

Finding the parameters $A_i$ using the boundary conditions Eqs.~(\ref{PVIa}), (\ref{PVIb}) and (\ref{pviderivative}), the variance is given by
\begin{equation}
\chi_2(t)=\frac{1}{k_i}-\frac{\Delta k}{k_i k_f}\left(3s^2-2s^3\right)+A_4 t_{f}^4\left(s^2-2s^3+s^4\right),
\end{equation}
where $s=t/t_f$ is the reduced time and $\Delta k= k_f-k_i$. Substituting this in Eq.~(\ref{n2}), the function $k(t)$ must be
\begin{equation}\label{kt}
k(t)=k_i \frac{1+\frac{3}{t_f}\frac{\Delta k}{k_i}(s-s^2)-\epsilon\frac{k_f}{k_i}\frac{1}{t_f}(s-3s^2+2s^3)}{1-\frac{\Delta k}{k_f}(3s^2-2s^3)+\epsilon(s^2-2s^3+s^4)},
\end{equation}
where $\epsilon=A_4k_it_f^4$. 

%\dr{Reference~\cite{trizac1} also proposed a similar protocol and carried out experimental realizations, achieving a reduction of two orders of magnitude in the equilibration time.} \\
The stochastic work \cite{sekimoto,pelitibook} done on the system during the time interval from $t=0$ to $t$ is given by
\begin{equation}\label{def work}
W=\frac{1}{2}\int_0^{t}x(t')^2\frac{dk(t')}{dt'}dt'.
\end{equation}
Computing the average of this quantity and substituting Eq.~(\ref{kt}), we obtain
\begin{align}
\expval{W}&=\frac{1}{2}\ln\left(\frac{k_f}{k_i}\right)+\frac{1}{4t_f}\frac{k_f}{k_i}\eta,
\nonumber\\
&=\Delta F + \expval{W_{\text{irr}}}
\end{align}
where $\eta$ has the expression
\begin{equation}
\eta=\int_0^1\frac{\left[\frac{-6\Delta k}{k_f}(s-s^2)+\epsilon (2s-6s^2+4s^3)\right]^2}{1-\frac{\Delta k}{k_f}(3s^2-2s^3)+\epsilon(s^2-2s^3+s^4)}ds,
\end{equation}
$\Delta F=\frac{1}{2}\ln(k_f/k_i)$ is the free energy difference between the final and initial states, and $\expval{W_{\text{irr}}}$ is the irreversible work.
The average work can be interpreted as a function of $\epsilon$ and numerical methods can be used to determine the value of $\epsilon$ that minimizes the average work. It is worth noting that the irreversible work $\expval{W_{\text{irr}}}$ exhibits an inverse relationship with the duration of the protocol $t_f$. This inverse relation is consistent with our expectation for a process aimed at accelerating equilibration. The external control exerted on the system requires additional work to achieve equilibrium within a shorter timeframe; however, this additional work should be optimized to achieve an efficient process.

\begin{figure}[t]
	\begin{center}
		\includegraphics[scale=0.6]{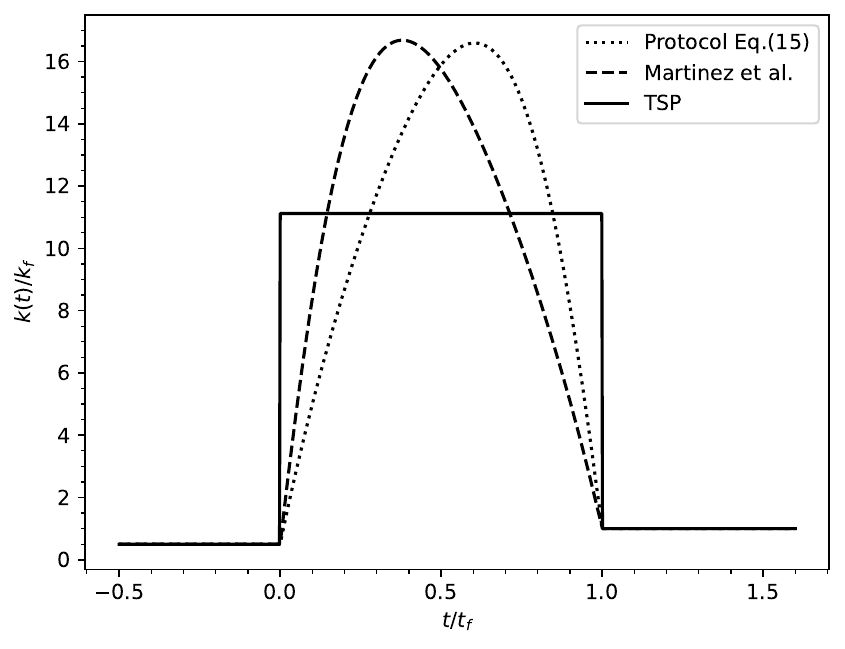}
	\end{center}
	\caption{\label{fig:stif}\small Time evolution of three different protocols for controlling the stiffness parameter $k$ in a harmonic oscillator. The protocols are: (a) the protocol given by Eq.~(\ref{kt}), (b) the protocol proposed by Martinez et al.~(Ref.~\cite{trizac1}), and (c) the TSP protocol defined in Eq.~(\ref{eq:TSP}). The common parameters for all protocols are $k_i=1/2$, $k_f=1$, and $t_f=1/30$. The value of $k_m$ for the TSP protocol is obtained by solving Eq.~(\ref{consistency}).} 
 %The maximum value of $k$ for each protocol is larger than $k_i$ and $k_f$, in order to reduce the equilibration time of the system.}

 %Plots of the protocols Eq.~(\ref{kt}), Martinez et al.\cite{trizac1}, and TSP Eq.~(\ref{eq:TSP}) with respect to time. The plot parameters are $k_i=1/2$, $k_f=1$, $t_f=1/30$ and $k_m$ is the solution of Eq.~(\ref{consistency}). The maximum of each protocol is greater than the initial and final values of $k$. This large value is necessary to reduce the equilibration time.}
\end{figure}

Since the choice of the functional form of $\chi_2$ is arbitrary, there exists an infinite number of solutions to the fast equilibration problem. Once we propose a form for $\chi_2(t)$, it is possible to find the corresponding protocol $k(t)$ using Eq.~(\ref{n2}). For example, in Ref.~\cite{trizac1}, a polynomial of degree 3 was proposed for the inverse of $\chi_2(t)$, and the corresponding experimental realization using optical tweezers was performed. The experimental results proved that their protocol shortened the relaxation time by two orders of magnitude. In order to make a comparison, we use the same values reported in Ref.~\cite{trizac1}, which in dedimensionalized units are $k_i=k_f/2=1/2$ and $t_f=1/30$, see Fig.~\ref{fig:stif}. The average work for protocol $k(t)$ defined in Eq.~(\ref{kt}), optimizing $\epsilon$ to minimize the average work yields $\expval{W}=6.52$, whereas the result published in Ref.~\cite{trizac1} yields $\expval{W}=6.71$. 
%It is worth noting that the optimal work protocol can be obtained using Pontryagin's principle (see Eq.~(\ref{eq:optimal})), with a result of $\expval{W}_{\text{opt}}=5.49$ \cite{trizac1, trizac3, schmiedlEfficiencyMaximumPower2007, Pontryagin64, Bechhoefer_control21}.

%The figure \ref{fig:stif} shows the difference between our protocol~(\ref{kt}) and the protocol reported at \cite{trizac1}.

%Once we propose a form for $\chi_2(t)$, it is possible to find the corresponding protocol $k(t)$ using Eq.~(\ref{n2}). Clearly, there are infinite protocols with different characteristics.
For example, for a variance of the form
\begin{equation}
    \chi_2(t)= \frac{1}{k_i} (1 + c t)^2,
\end{equation}
the protocol is
\begin{equation}\label{eq:optimal}
k_{\text{op}}(t) = \frac{k_i}{(1+c t)^2} - \frac{c}{1+c t},
\end{equation}
where $c = \frac{1}{t_f} (\sqrt{k_i/k_f}-1)$ and the time interval $t\in (0,t_f)$. This protocol has the peculiarity of minimizing average work, as shown in Refs.~\cite{schmiedlEfficiencyMaximumPower2007,trizac3}. Another example is a linear relationship for the variance $\chi_2(t)$ that satisfies the boundary conditions in Eqs.~(\ref{PVIa})-(\ref{PVIb})
\begin{equation}
    \chi_2(t)= \frac{1}{k_i}+\left(\frac{1}{k_f}-\frac{1}{k_i}\right)\frac{t}{t_f},
\end{equation}
whose protocol $k_L(t)$ is of the form for $t\in (0,t_f)$
\begin{equation}\label{eq:linear}
    k_{L}(t)=\frac{2-\left(\frac{1}{k_f}-\frac{1}{k_i}\right)\frac{1}{t_f}}{2\left(\frac{1}{k_i}+(\frac{1}{k_f}-\frac{1}{k_i})\frac{t}{t_f}\right)}.
\end{equation}
Both protocols are discontinuous at $t=0$ and $t=t_f$ and they will be called optimal and linear, respectively. 
Studying the energetics \cite{sekimoto} of these protocols, as well as almost all other protocols, can only be achieved through numerical simulations \cite{Speck}. This raises the question of whether it is possible to design a protocol for obtaining analytical solutions for the PDF of work, heat, and produced entropy. The following section addresses this question.

\section{Two-step protocol}\label{sec:TSP}

In this section, we propose a new fast thermalization protocol that provides analytical solutions for the PDF of the relevant thermodynamic quantities. After careful examination of Fig.~\ref{fig:stif}, it is evident that the stiffness must be significantly greater than the initial and final values of $k$. This is due to the fact that to accelerate the equilibration process, the stiffness must be increased to reduce the characteristic timescale evolution of the system. Therefore, it is reasonable to conclude that large stiffness values play a critical role in achieving a shortcut to adiabaticity.

Considering this, we propose a two-step protocol (TSP) defined as follows

\begin{align}\label{eq:TSP}
    k(t)= 
    \begin{cases}
            k_i, &  t \leq 0\\
            k_m, &  0<t<t_f\\
            k_f, & t\geq t_f.  
    \end{cases}
\end{align}
As we will see, this toy model captures the essential features of general protocols. Using this stiffness, the variance in the time interval $0\leq t \leq t_f$ was obtained by solving Eq.~(\ref{n2}) with the initial condition of Eq.~(\ref{PVIa})
\begin{equation}\label{chi}
\chi_2(t)=\sigma_X(t)^2=\frac{1}{ k_m}+\left(\frac{1}{ k_i}-\frac{1}{ k_m}\right)e^{-2k_mt}.
\end{equation}
Using this variance, the expression for the probability distribution of the position at any time ($0\leq t\leq t_f$) can be written as
\begin{equation}\label{position pdf}
P_X(x_t,t)=\frac{1}{\sqrt{2\pi}\sigma_X(t)}\exp\left(-\frac{x_t^2}{2\sigma_{X}^2(t)}\right).
\end{equation}
By forcing the system to arrive at a final equilibrium state at $t=t_f$, the variance $\chi_2(t_f)$ must be $1/k_f$. This leads to a self-consistency equation for $k_m$ given by
\begin{equation}\label{consistency}
\exp(-2k_m t_f)=\frac{\displaystyle\frac{1}{k_f}-\frac{1}{k_m}}{\displaystyle\frac{1}{k_i}-\frac{1}{k_m}}.
\end{equation}
This equation is transcendental and can be solved numerically to determine the value of $k_m$. Utilizing this value, the TSP is a control protocol that joins the equilibrium state at $k_i$ with the equilibrium state at $k_f$ in a finite time $t_f$. The Eq.~(\ref{consistency}) is not symmetric under the time inversion of the protocol, i.e, if the system evolves from $k_f$ to $k_i$, the value of $k_m$ changes. Therefore, the system does not follow the same evolution in the reverse direction of time. This fact is related to the produced entropy, which is discussed later in this paper.

This approach differs from that in the previous section and previous works Refs.~\cite{trizac1, bayatiDiffusiophoresisDrivenColloidal2021}, in which a specific form of the protocol is proposed and the corresponding variance follows from Eq.~(\ref{n2}). Thus, this process is referred to as direct engineering.

Although Eq.~(\ref{consistency}) cannot be solved analytically, we can derive the asymptotic behavior of $k_m$ for large and small values of target final time $t_f$. If $t_f\gg 1$, $\exp(-2k_m t_f)\to 0$. From the right-hand side of Eq.~(\ref{consistency}), we find $k_m=k_f=1$. This is expected, as if we have an infinite amount of time to allow the system to equilibrate ($t_f\to\infty$), we can fix $k_m=k_f$ and wait for the system to reach equilibrium at this new stiffness value.

More interesting and pertinent is the behavior when $t_f\to 0$. In this limit, $k_m\to\infty$, which means that the shorter the target time, the larger the stiffness. At this limit, the right-hand side of Eq.~(\ref{consistency}) converges to a finite value $k_i/k_f$. Therefore, we deduce that $k_m t_f$ must remain finite, and we get
\begin{equation}
\label{eq:km_tf_short}
k_m \sim \frac{1}{2t_f} \ln\frac{k_f}{k_i}.
\end{equation}
It is observed that $k_m$ is inversely proportional to $t_f$. This quantifies the previous observation that, to achieve fast relaxation to equilibrium, it is necessary to significantly increase the stiffness. Since the typical relaxation time at a fixed stiffness $k_m$ is of order $1/k_m$, the TSP matches the transient relaxation time of order $1/k_m$ to the target duration $t_f$ of the accelerated protocol. The specific relationship is given by Eq.~(\ref{eq:km_tf_short}).

We can compute the corrections to Eq.~(\ref{eq:km_tf_short}) by expanding the right-hand side of Eq.~(\ref{consistency}) in powers of $1/k_m$. For example, the next-to-leading-order term is given by
\begin{equation}
\label{eq:km_tf_short2}
k_m t_f = \frac{1}{2}\ln\frac{k_f}{k_i}
+ t_f \frac{k_f-k_i}{\ln(k_f/k_i)} +O(t_f^2).
\end{equation}
It turns out that the leading order Eq.~(\ref{eq:km_tf_short}) is universal for more general protocols if we substitute $k_m$ by the average value of $k(t)$ over the protocol duration 
\begin{equation}
    k_m=\frac{1}{t_f} \int_0^{t_f} k(t)\,dt.
\end{equation}

\begin{figure}[H]
	\begin{center}
		\includegraphics[scale=0.6]{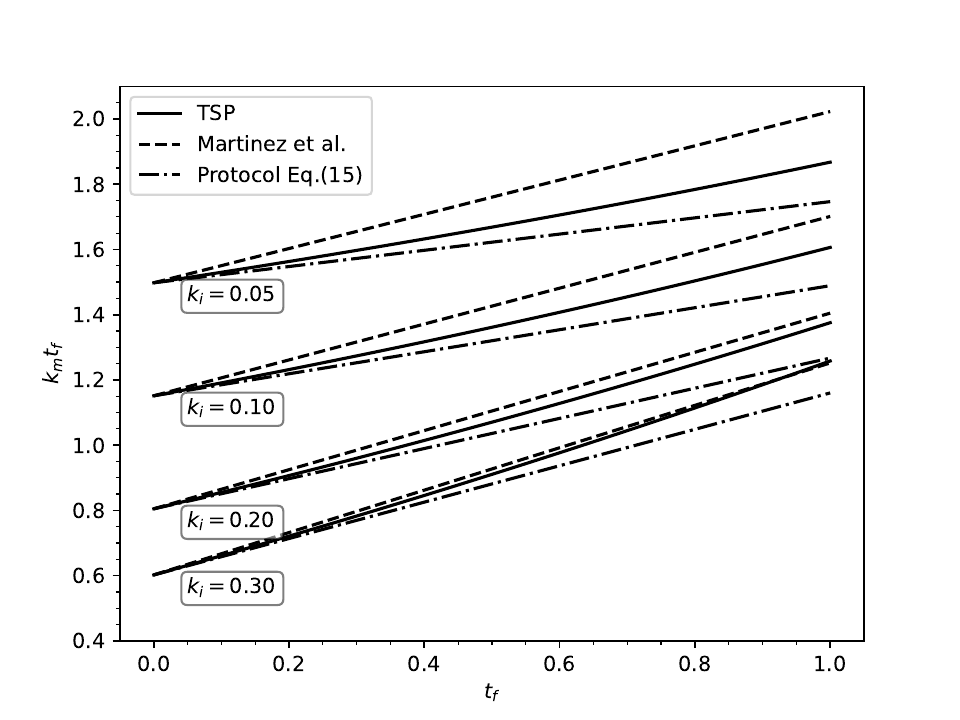}
	\end{center}
	\caption{\label{fig:kmtf_protocols}
 The average value $k_m$ of the stiffness multiplied by $t_f$ is plotted for three protocols: the Two-Step Protocol (TSP) given by Eq.~(\ref{eq:TSP}), the protocol proposed by Martinez et al.~(Ref.~\cite{trizac1}), and the protocol $k(t)$ given by Eq.~(\ref{kt}). It can be observed that at short target times $t_f$, all protocols exhibit the same asymptotic behavior, as predicted by Eq.~(\ref{eq:km_tf_short}).}
\end{figure}

Indeed, when Eq.~(\ref{n2}) is used for a general protocol, we find that the average of $k(t)$ is 
\begin{equation}
    k_m = \frac{1}{2t_f}\ln\frac{\chi_2(0)}{\chi_2(t_f)}
    + \int_0^{t_f} \frac{1}{\chi_2(t)}\frac{dt}{t_f}
    .
\end{equation}
Since $\chi_2$ is a continuous function of $t$, applying the mean value theorem 
and using the initial and final conditions Eqs.~(\ref{PVIa})-(\ref{PVIb}) leads to
\begin{equation}
    k_m = \frac{1}{2t_f}\ln\frac{k_f}{k_i}
    + \frac{1}{\chi_2(t^*)}
\end{equation}
for some $t^*\in[0,t_f]$. 
Figure~\ref{fig:kmtf_protocols} illustrates this by showing a plot of $k_m t_f$ as a function of $t_f$ for different values of $k_i$ and the ESE protocols discussed in the previous section and Ref.~\cite{trizac1}. As $t_f\to 0$, all the protocols converge to the same value given by Eq.~(\ref{eq:km_tf_short}). However, the next-to-leading corrections differ from one protocol to another.

%The parameter $k_m$ is a constant value that is fixed only if we assume a final equilibrium situation at time $t_f$. At present, $k_m$ remains general until further clarification is provided.

\subsection{Work distribution for two-step protocol}\label{sub_sec:work pdf}
Since the TSP seems to be analytically tractable and exhibits some general features of more complex ESE protocols, we now proceed to use it as a basis for understanding the stochastic energetics of the fast relaxation protocols. More specifically, in this section, we want to calculate the work probability distribution.
Using Eq.~(\ref{def work}), we obtain the work $W_D$ done on the system during the process
\begin{align}\label{eq:work done}
    W_D= \left\{ \begin{array}{lcc}
              \frac{1}{2}(k_m-k_i)x_i^2, &  \quad 0< t < t_f\\
             \\\frac{1}{2}(k_m-k_i)x_i^2+\frac{1}{2}(k_f-k_m)x_f^2,&  \quad  t>t_f.
             \end{array}
   \right. %para finalizar (delimitador invisible con el punto).
\end{align}
Since the initial and final positions are stochastic variables with normal distributions, the work done is also a stochastic variable. To compute the work distribution function, we need to consider the following two cases.
%and using the boundary conditions \ref{PVI}, the average work after the two steps is \footnote{$\expval{W}_{eq}$ is the average work assuming a final equilibrium situation.}
%\begin{equation}
%    \beta\expval{W}_{eq} =k_m(\frac{1}{k_i}-\frac{1}{k_f}) 
%\end{equation}
\subsubsection{For $0<t<t_f$}
To study this time interval, we need to determine the position probabilities at two points described by $(x_i, t_i=0)$ and $(x_t, t)$. Therefore, the joint probability distribution (JPD) can be expressed as follows
\begin{equation}\label{two points pdf}
P_2(x_i,t_i;x_t,t)=P(x_t,t|x_i,t_i)P(x_i,t_i).
\end{equation}
Computing $P_2$ is a straightforward task; $P(x_t,t|x_i,t_i)$ follows an Ornstein-Uhlenbeck process \cite{gardiner}
\begin{equation}
\label{eq:OH-km}
    P(x_t,t_t|x_i,t_i)=\frac{1}{\sqrt{2\pi \sigma_m(t)^2}}\exp{-\frac{(x_t-x_ie^{-k_mt})^2}{2\sigma_m(t)^2}},
\end{equation}
where 
\begin{equation}\label{sigma m}
    \sigma_m(t)^2 =\frac{1-e^{-2k_mt} }{ k_m }.
\end{equation}
The initial Gaussian probability distribution is
\begin{equation}\label{pdf initial}
    P(x_i,t_i)=\frac{1}{\sqrt{2\pi}\sigma_i}\exp{-\frac{x_i^2}{2\sigma_i^2}}
\end{equation}
with $\sigma_i^2=1/k_i$ (Eq.~(\ref{position pdf}) at $t=0$).

Considering there exists an infinite number of paths connecting the points $(x_i, t_i)$ and $(x_t, t)$ to compute the probability density function of the work, we adopt the following approach. We focus solely on the paths that yield a specific amount of work $W_D$, regardless of the initial and final points. Hence, the PDF of the work can be expressed as follows

%\begin{equation}\label{pdf work}
%P(W,t)=\int_{-\infty}^{\infty}dx_i dx_t \delta(W-W_D)P_2(x_i,t_i;x_t,t)
%\end{equation}

\begin{equation}\label{work}
  P_{W_D}(W,t)=\int_{-\infty}^{\infty}dx_i \int_{-\infty}^{\infty} dx_t \, \delta(W-W_D)P_2(x_i,t_i;x_t,t).
\end{equation}
Since the work $W_D$ is independent of $x_t$ for $t < t_f$, the work probability is stationary, and the previous equation simplifies to
\begin{equation}
    \label{eq:PWtfinf}
P_{W_D}(W, t) = \int_{-\infty}^{\infty} dx_i \delta\left(W - \frac{(k_m - k_i) x_i^2}{2}\right) P(x_i, t_i).
\end{equation}
If $W/(k_m - k_i) < 0$, the argument of the Dirac delta distribution never vanishes, resulting in $P(W, t) = 0$. However, when $W/(k_m - k_i) \geq 0$, there are two roots to the equation $W - (k_m - k_i) x_i^2/2 = 0$, yielding:
\begin{equation}
P_{W_D}(W, t) = \sqrt{\frac{k_i}{\pi (k_m - k_i) W}} \exp\left(-\frac{k_i}{k_m - k_i} W\right).
\end{equation}
This result can also be obtained using the methods described in Refs.~\cite{Speck, pelitibook}.
Returning to Eq.~(\ref{eq:PWtfinf}), one can use the Fourier representation of the Dirac distribution
\begin{equation}
\label{eq:deltaWtinftf}
    \delta(W-(k_m-k_i)x_i^2/2) = \int_{-\infty}^{+\infty} e^{iz(W-(k_m-k_i)x_i^2/2)}
    \,\frac{dz}{2\pi}
\end{equation}
to compute the characteristic function of $P_{W_D}(W,t)$. Inserting Eq.~(\ref{eq:deltaWtinftf}) into Eq.~(\ref{eq:PWtfinf}), the resulting Gaussian integral over $x_i$ can be performed to obtain 
\begin{equation}\label{characterictic W1}
    P_{W_D}(W,t)=\frac{1}{2\pi}\int_{-\infty}^{\infty}dz\hat{P}_W(z)e^{i W z},
\end{equation}
where the characteristic function is
\begin{equation}\label{linear}
    \hat{P}_{W_D}(z)=\frac{1}{\sqrt{1+ia_1 z}}
\end{equation}
with $a_1=(k_m-k_i)/k_i$. Using the expansions
%\dr{PREVIAMENTE LLAMASTE A LOS CUMULANTES DE OTRA MANERA.}
\begin{align}
    \ln \hat{P}_{W_D}(z) & = \sum_{n=1}^\infty 
    \frac{(-iz)^n}{n!} c_n 
    \nonumber\\
    &= \sum_{n=1}^\infty \frac{(-ia_1 z)^n}{2n},
\end{align}
we can obtain all the cumulants $c_n$ of $W$
\begin{equation}
    c_n = \frac{1}{2}(n-1)! a_1^n.
\end{equation}
In particular, the average work and variance for this time interval are
\begin{eqnarray}\label{average work and variance t lower}
  \expval{W}&=&\frac{k_m-k_i}{2k_i} ,\\
  \sigma_W^{2}&=&\frac{(k_m-k_i)^2}{2k_i^2}.
\end{eqnarray}
Note that $\sigma_W^{2}=2\expval{W}^2$. From the characteristic function, one can check that the work distribution satisfies the Jarzynski equality~\cite{Jar97} as it should
\begin{equation}
    \left<e^{-W}\right> = \hat{P}_{W_D}(i)
    = \sqrt{\frac{k_i}{k_m}} = e^{-\Delta F},
\end{equation}
where $\Delta F=\frac{1}{2}\ln(k_m/k_i)$ is the free energy difference for the oscillator with final stiffness  $k_m$ (in the time interval $0<t<t_f$) and initial stiffness $k_i$.

\subsubsection{For $t\geq t_f$}
For the second case of the computation of work PDF is necessary to consider the three-point joint probability distribution
\begin{equation}
    P_3(x_i,t_i;x_f,t_f;x,t)=P(x,t|x_f,t_f)P_{2}(x_f,t_f;x_i,t_i).
\end{equation}
Again, the work distribution is stationary; hence, we only need the two-point JPD
\begin{align}
    P_2(x_i,t_i;x_f,t_f)&=\frac{1}{2\pi \sigma_i\sigma_m(t)}\exp{-\frac{(x_f-x_ie^{-k_mt_f})^2}{2\sigma_m(t_f)^2}
    }
    \nonumber\\
    &\times \exp{-\frac{x_i^2}{2\sigma_i^2}}.
\end{align}
Here $\sigma_m(t)$ is given by Eq.~(\ref{sigma m}).
Inserting the two-point JPD in an expression similar to Eq.~(\ref{work}) and using the Fourier representation of the Dirac distribution, we obtain a Gaussian integral in two variables that can easily be computed leading to a characteristic function of the form
\begin{equation}\label{cuadratic}
    \hat{P}_{W_D}(z)=\frac{1}{\sqrt{1+a_1 z+a_2 z^2}}
\end{equation}
where the coefficients are given by 
\begin{eqnarray}
  a_1&=&\frac{i(k_f-k_i)}{k_i}+\frac{i(k_f-k_m)(k_i-k_m)\sigma_m^2(t_f)}{k_i}, \\ 
  a_2&=&
\frac{i(k_f-k_m)(k_i-k_m)\sigma_m^2(t_f)}{k_i}.
\end{eqnarray}
Substituting the consistency equation Eq.~(\ref{consistency}), these expressions can be simplified to
\begin{eqnarray}
  a_1&=&\frac{i(k_f-k_i)k_m}{k_ik_f}, \\ 
  a_2&=&\frac{(k_f-k_i)(k_m-k_f)}{k_fk_i}.
\end{eqnarray}
Clearly, $P_{W_D}(W)$ is normalized ($\hat{P}_W(0)=1$) and satisfies the Jarzynski equality~\cite{Jar97}
\begin{equation}
    \expval{e^{- W}}=
    \hat{P}_{W_D}(i)=
    e^{- \Delta F},
\end{equation}
where now $\Delta F=\frac{1}{2}\ln(k_f/k_i)$ is the Helmholtz free energy difference between the two equilibrium states with stiffness $k_f$ and $k_i$.
%\begin{equation}\label{pdf work integral}
%    P(W)=\frac{\beta}{2\pi\sigma_i\sigma_m}\int_{-\infty}^{\infty}dz \frac{e^{i\beta W z}}{\sqrt{Det(A)}}
%\end{equation}
%where the matrix $A$ is given by
%\[
%A =
%\begin{pmatrix}
%i(k_m-k_i)\beta z+\frac{1}{\sigma_i^2}+\frac{e^{-2t_f/\tau_m}}{\sigma_m^2} & -\frac{e^{-t_f/\tau_m}}{\sigma_m^2} \\
%-\frac{e^{-t_f/\tau_m}}{\sigma_m^2} & i(k_f-k_m)\beta z+\frac{1}{\sigma_m^2}
%\end{pmatrix}
%\]
From the characteristic function, we can determine any moments of work distribution, including the average work and variance
\begin{eqnarray}
    \expval{W} &=& \frac{k_m}{2}\left(\frac{1}{k_i}-\frac{1}{k_f}\right), \\
    \sigma_W^2 &=& 2\expval{W}^2 + \frac{2(k_m - k_f)\expval{W}}{k_m}.
\end{eqnarray}
It is worth noting that $\sigma_W^2 > 2\expval{W}^2$ holds true when $(k_m - k_f)\expval{W} > 0$. 
%Figure~\ref{fig:work_metrics} shows the average work and its standard deviation as function of time.
%\begin{figure}[t]
%	\begin{center}
%		\includegraphics[scale=0.6]{work_metrics}
%	\end{center}
%	\caption{
% \label{fig:work_metrics} \small {The standard deviation and work average vs time for the TSP protocol whose parameters are $k_i=k_f/2=1/2$ and $t_f=1/30$.}}
%\end{figure}

The exact computation of the inverse Fourier transform in Eq.~(\ref{characterictic W1}) can be achieved for the characteristic function given in Eq.~(\ref{cuadratic}). To accomplish this, it is imperative to use the vertex form of a quadratic polynomial. Subsequently, an elementary change of variable results in an integral representation of the modified Bessel function of the second kind of order zero $K_0$. The final expression is as follows

\begin{widetext}
\begin{align}\label{eq:pdf work}
    P_{W_D}(W)&=\frac{1}{\pi\sqrt{\sigma_W^2-2\expval{W}^2}}\exp{\frac{\expval{W}}{\sigma_W^2-2\expval{W}^2}W}K_0\left(\frac{\sqrt{\sigma_W^2-\expval{W}^2}}{\sigma_W^2-2\expval{W}^2}\abs{W}\right).
    %\\
    %&
    %\gt{
    %=
    %\frac{1}{\pi\sqrt{
    %2\expval{W}(k_m-k_f)/k_m
    %}}
    %\exp{
    %\frac{k_m}{2(k_m-k_f)}
    %W}
    %K_0\left(\frac{k_m\sqrt{\sigma_W^2-\expval{W}^2}}{2(k_f-k_m)\expval{W}}\abs{W}%\right)
    %}
\end{align}
%\gt{XXX: \em Decidir cual de las dos formas dejamos. Propongo la segunda ya que el factor exponencial toma una forma más explícita en función de los $k$. Además tiene gran similitud con la del calor.}
\end{widetext}

\begin{figure}[t]
	\begin{center}
		\includegraphics[scale=0.6]{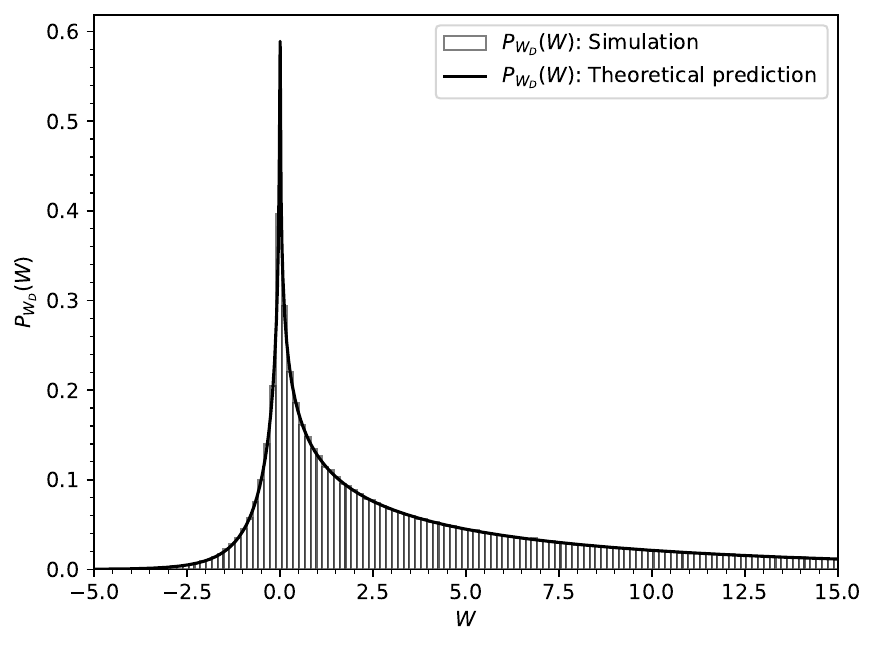}
	\end{center}
	\caption{
 \label{fig:work_pdf}\small{
 Comparison between the theoretical and numerical probability density functions (PDFs) of the work done on a harmonic oscillator after the stiffness parameter $k$ is changed according to TSP. The solid curve is the theoretical prediction for $t>t_f$ given by Eq.~(\ref{eq:pdf work}). The histogram is obtained by simulating the Langevin dynamics of the system with $10^6$ realizations. The parameters for the simulations are $k_i=1/2$, $k_f=1$, and $t_f=1/30$. }}
 \end{figure}

\begin{figure}[t]
	\begin{center}
		\includegraphics[scale=0.6]{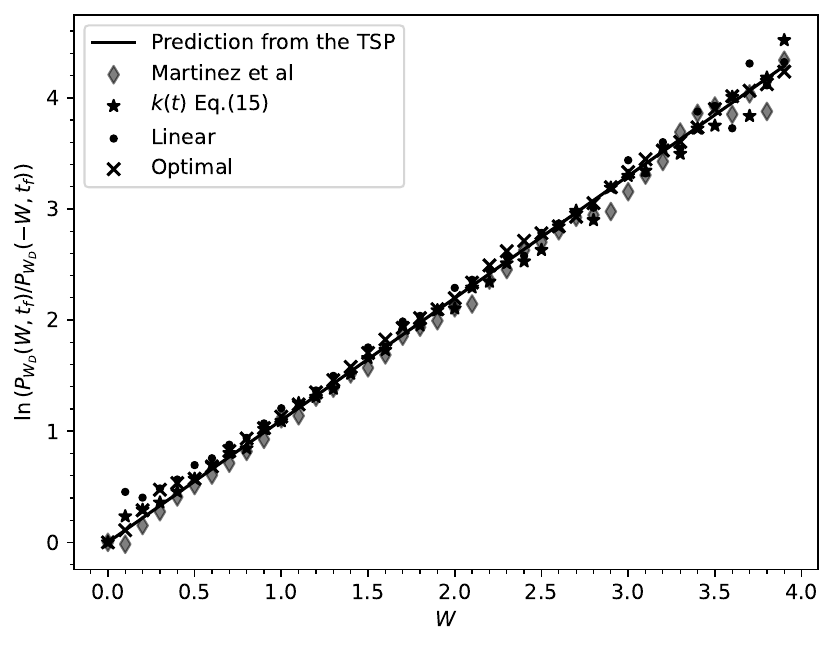}
	\end{center}
	\caption{
 \label{fig:hypothesis_W}\small{
 We tested the TSP relation, Eq.~(\ref{eq:PWP-W}), for four different protocols: the protocol by Martinez et al. Ref.~\cite{trizac1}, the protocol in Eq.~(\ref{kt}), and the linear and optimal protocols in Eqs.~(\ref{eq:linear}) and ~(\ref{eq:optimal}), respectively. We used $10^6$ simulations with $k_i=1/2$, $k_f=1$, and $t_f = 1/30$.}}
 %Test of the relation, Eq.~(\ref{eq:PWP-W}), implies that $P(W)W \propto 1/t_f$. The work PDFs are from $10^6$ simulations with $k_i=1/2$, $k_f=1$, and $t_f = 1/30$.}}
 
 %The protocol given by Martinez et al.~\cite{trizac1}, the protocol described in Eq.~(\ref{kt}), and the linear (\ref{eq:linear}) and optimal protocol (\ref{eq:optimal}) exhibit a relation predicted by the TSP in Eq.~(\ref{eq:PWP-W}). We have performed $10^6$ simulations using $k_i=1/2$, $k_f=1$, and $t_f = 1/30$.}}
\end{figure}

The comparison between the above work probability distribution and simulations is depicted in Fig.~\ref{fig:work_pdf}. The simulations have been performed with an open source code developed by us, available in Ref.~\cite{langesim}.
The work PDF Eq.~(\ref{eq:pdf work}) exhibits an intriguing structure characterized by the product of two components. The first component is an exponential function of $W$, introducing an asymmetry in the probabilities of obtaining positive or negative work values. The second component is a symmetric function of $W$, specifically the Bessel $K_0$ function, with an argument proportional to the absolute value of $W$. Since the overall process is a compression ($k_f>k_i$), the average work done on the particle is positive. Nevertheless, there are rare events in which $W<0$ but those are less frequent than the ones where $W>0$. Accordingly, the tail of $P_{W_D}(W)$ for $W>0$ is larger than the one for $W<0$. Mathematically, this asymmetry factor can be quantified by the ratio $P_{W_D}(W)/P_{W_D}(-W)$. From Eq.~(\ref{eq:pdf work}), we have
\begin{equation}
\label{eq:PWP-W}
    \frac{P_{W_D}(W)}{P_{W_D}(-W)}=\exp{\frac{2\expval{W}}{\sigma_W^2-2\expval{W}^2}W}=
    \exp{\frac{k_m}{k_m-k_f}W}
    \,.
\end{equation}
This equation is a consequence of the symmetry relation for the characteristic function
\begin{equation}
    \hat{P}_{W_D}(z) = \hat{P}_{W_D}\left(-z-i\frac{k_m}{k_m-k_f}\right).
\end{equation}
%In Fig.~\ref{fig:hypothesis_W} the relationship between Eq.~(\ref{eq:PWP-W}) for the TSP protocol is outlined,
Figure~\ref{fig:hypothesis_W} shows the relationship outlined in Eq.~(\ref{eq:PWP-W}) for the TSP protocol,  alongside the results of the other protocols obtained by simulations. From this figure, we observe that the behavior of the ratio $P_{W_D}(W)/P_{W_D}(-W)$ is the similar for all protocols when $t_f\ll 1$ (the range of interest for fast thermalization protocols). This numerical evidence leads us to conjecture that this is a general property. An analytical proof of this fact for an arbitrary protocol remains as an open question. 

%The equation ~(\ref{eq:PWP-W}) has been plotted in~\ref{fig:hypothesis_W} alonside with the numerical results of the protocols cited in this paper.

 %The curve is the theoretical prediction for $t>t_f$ given by (\ref{eq:pdf work}) and the histogram is built simulating the Langevin dynamic of the system under study. This histogram is for $10^6$ simulations with parameters $k_i=1/2$, $k_f=1$, and $t_f=1/30$. }}
%\end{figure}\\

\begin{figure}[t]
	\begin{center}
		\includegraphics[scale=0.6]{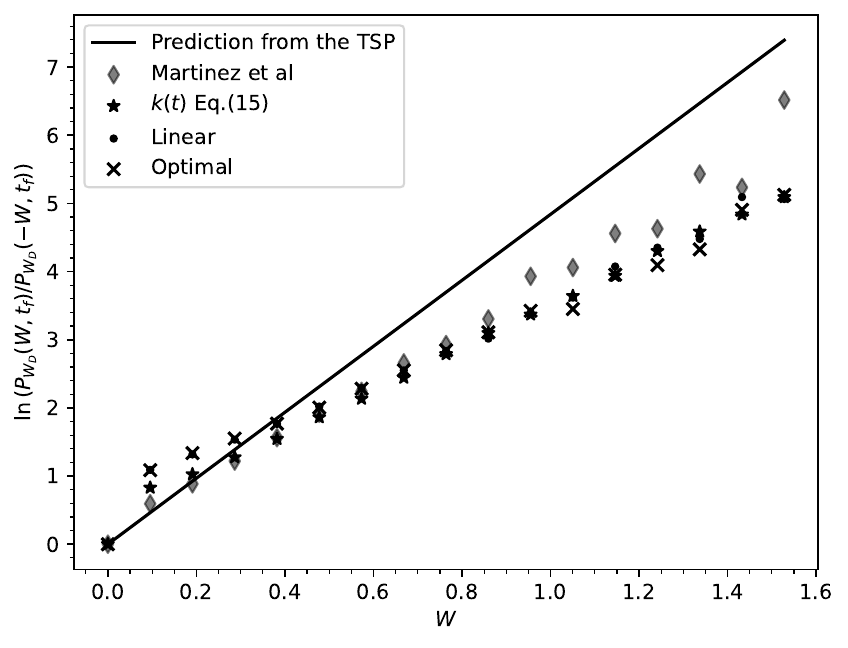}
	\end{center}
	\caption{
 \label{fig:scatter_wtf}\small{The behavior predicted by Eq.~(\ref{eq:PWP-W}) present a deviation for $t_f\sim 1$. However, we can observe a linear tendency.
 This plot has been made for $10^6$ simulations with parameters $t_f=0.7$, $k_i=1/2$, and $k_f=1$.}}
\end{figure}

%that the relation given by Eq.~(\ref{eq:PWP-W}) is \dr{independent} of the protocol considered that achieve fast thermal equilibration in a short time, where $t_f \ll \taurel = 1$. 
However, the behavior predicted by Eq.~(\ref{eq:PWP-W}) breaks down at times $t_f \sim \taurel=1$. The other protocols exhibit slight deviations from the predicted behavior of the TSP protocol but still follow a linear behavior (in semi-log scale), as shown in Fig.~\ref{fig:scatter_wtf}.
\\
The tails of the work distribution Eq.~(\ref{eq:pdf work}) can be obtained from the asymptotic behavior of the Bessel function $K_0$, they are 
\begin{align}
    P_{W_D}(W) \sim\left\{ \begin{array}{lcc}
              \frac{1}{\sqrt{2\pi\sigma_r W}}\exp{\frac{-1}{\sigma_r+\expval{W}}W} &  \quad W\rightarrow\infty, \\
             \frac{1}{\sqrt{-2\pi\sigma_r W}}\exp{\frac{1}{\sigma_r-\expval{W}}W}&  \quad W\rightarrow -\infty  .
             \end{array}
   \right. %para finalizar (delimitador invisible con el punto).
\end{align}
where $\sigma_r^2=\sigma_W^2-\expval{W}^2$. 

%\dr{The simulations have been performed with an open code available in ~\cite{langesim}.}
\subsubsection{Reverse protocol}

Eq.~(\ref{eq:PWP-W}) is different from the Crook's fluctuation theorem Refs.~\cite{Crooks1998, crooks_entropy_1999}, which we review and test for our model in this section. Previously, we computed the probability distribution connecting the equilibrium states for $k_i$ and $k_f$, through a suitable choice of $k_m$ as given by Eq.~(\ref{consistency}). However, our results are more general than this specific case. The probability density function of work retains the same expressions as before for arbitrary $k_m$, except that the average work and its variance will change. In this general setup, the expressions for $t < t_f$, remain unchanged, see Eq.~(\ref{average work and variance t lower}). The expressions for $t\geq t_f$ are as follows
\begin{eqnarray}\label{metrics general}
\expval{W}&=&\frac{k_f-k_i}{2k_i}+\frac{(k_m-k_f)(k_m-k_i)}{2k_i}\sigma_m^2(t_f),
\\  
\label{eq:metrics-gen-sigma}
\sigma_W^2&=& 2\expval{W}^2+\frac{(k_m-k_f)(k_m-k_i)}{k_i}\sigma_m^2(t_f).
\end{eqnarray}

Now, let us consider a time-reversed protocol defined as
\begin{align}\label{eq:RTSP}
    k^R(t)= \begin{cases}
             k_f, &   t \leq 0\\
             k_m, &  0<t<t_f\\
             k_i, &   t\geq t_f.
             \end{cases}
\end{align}
In this time-reversed protocol, we do not change the value of $k_m$, it is the same as in the forward protocol. As a result, the system will {\em not} be at thermal equilibrium at $t_f$ since the value of $k_m$ has not been adjusted properly (to obtain thermal equilibrium at $t_f$ in a reversed protocol, the roles of $k_i$ and $k_f$ had to be interchanged in Eq.~(\ref{consistency})).

The Crooks relation can be verified by utilizing the general expression for $\expval{W}$ and $\sigma_W^2$ \cite{crooks_entropy_1999}
\begin{equation}\label{eq:crooks identity}
    \frac{P_{W_D}(W)}{P^R_{W_D}(-W)}=\exp{W-\Delta F},\quad \text{for any $t$}.
\end{equation}
%\vspace{0.3cm}
Here, $P^R_W(-W)$ represents the work probability distribution of the time reversed protocol. On the left-hand side of Eq.~(\ref{eq:crooks identity}), the cancellation of Bessel functions occurs due to the symmetry of the argument under the interchange of $k_i$ and $k_f$. The factor $\Delta F$ arises from the ratio of normalization factors in Eq.~(\ref{eq:pdf work}), while the appearance of $W$ stems from the exponential function (refer to Appendix~\ref{app:1} for more details).

\subsection{Heat distribution function for TSP}\label{sub_sec:heat pdf}
In this section, we move to the study of another relevant thermodynamics quantity: heat.
For a Brownian particle confined in a harmonic potential, the heat that enters to the system is expressed as \cite{sekimoto}
\begin{equation}
Q_D(t)=\int_0^t k(t')x(t')\dot{x}(t')dt'
.
\end{equation}
Using the first law of thermodynamics, the heat can also be computed from $Q_D(t)= U(t)-U(0)-W_D(t)$.
For the TSP, the stochastic heat is
\begin{align}\label{eq:heat done}
    Q_D(t)= \left\{ \begin{array}{lcc}
             \frac{1}{2}k_m(x_t^2-x_i^2), & \quad t \leq t_f\\
             \\\frac{1}{2}k_m(x_f^2-x_i^2)+\frac{1}{2}k_f(x_t^2-x_f^2),& \quad t>t_f.
             \end{array}
   \right. %para finalizar (delimitador invisible con el punto).
\end{align}
%Notably, the average heat after the two steps, assuming final equilibrium is reached at $t=t_f$ and utilizing the result from \ref{PVI}, can be calculated as:

%\begin{equation}\label{eq heat}
%\expval{Q}_{eq}=\frac{k_m}{2\beta}\left(\frac{1}{k_f}-\frac{1}{k_i}\right)
%\end{equation}
It is worth noting that the heat distribution is time-dependent because it involves $x_t$. This complication leads to a non-stationary heat probability density function. Similar to the work calculation, the analysis of heat needs to be performed in two distinct cases.
\subsubsection{For $0<t<t_f$}

\begin{figure}[t]
	\begin{center}
		\includegraphics[scale=0.6]{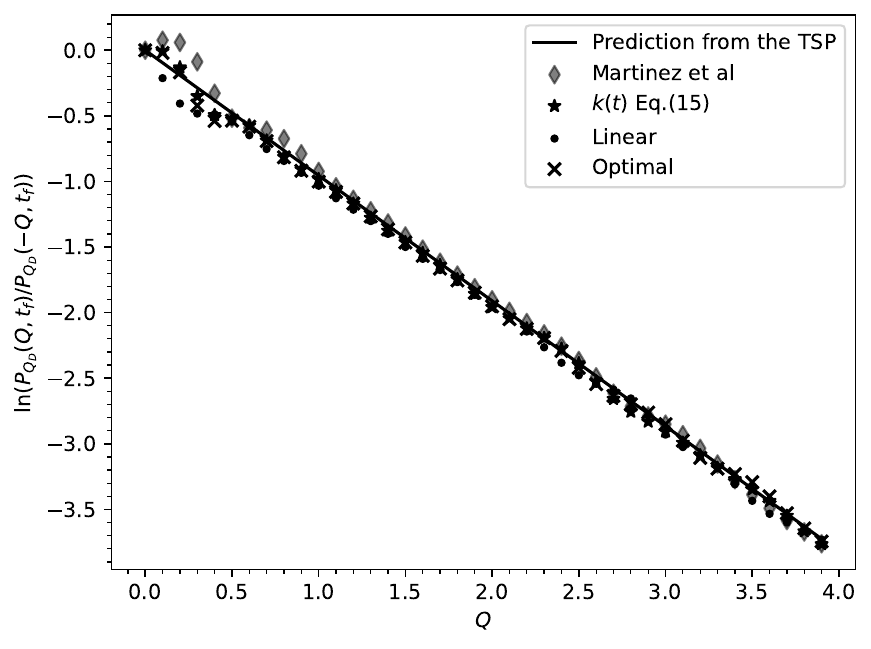}
	\end{center}
	\caption{
 \label{fig:hypothesis_Q}
 \small{ The relation established for the TSP protocol, as expressed in Eq.~(\ref{eq:PQP-Q}), is followed by more generalized protocols as the Martinez et al. Ref.~(\cite{trizac1}), the $k(t)$ protocol defined in Eq.~(\ref{kt}), the linear protocol Eq.~(\ref{eq:linear}), and the optimal protocol Eq.~(\ref{eq:optimal}). This behavior is observed under the condition $t_f\ll1$. The figure is constructed based on $10^6$ simulations for the parameters $k_i = 1/2$, $k_f=1$, and $t_f = 1/30$. }}
 
 %The protocol described by Martinez et al.~Ref~\cite{trizac1}, the protocol $k(t)$ described in Eq.~(\ref{kt}), the linear Eq.~(\ref{eq:linear}) and the optimal protocol Eq.~(\ref{eq:optimal}) follows the prediction given by the TSP Eq.~(\ref{eq:PQP-Q}) for $t_f\ll1$. This observation is based on $10^6$ simulations performed with  }}

\end{figure}

\begin{figure}[t]
	\begin{center}
		\includegraphics[scale=0.6]{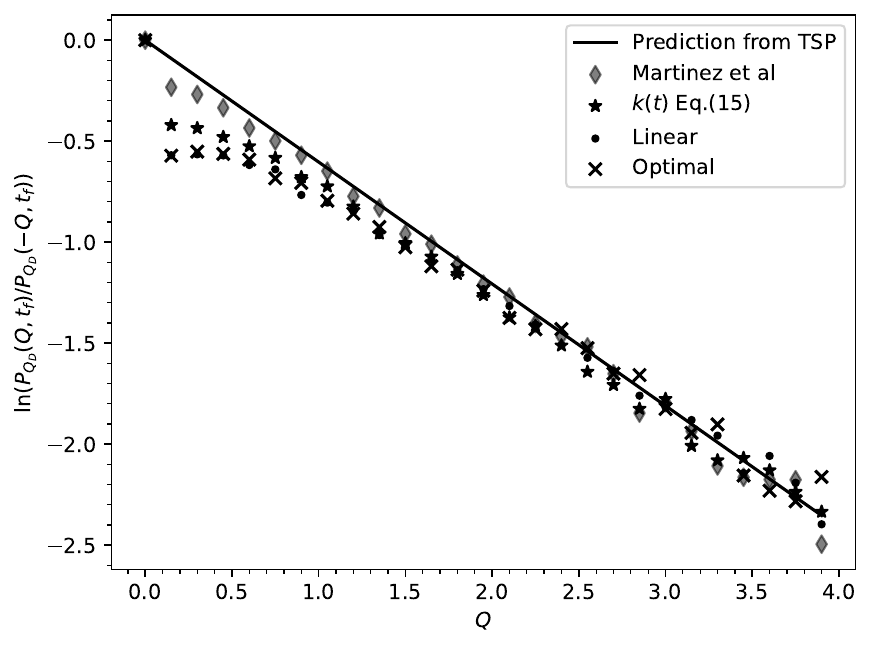}
	\end{center}
	\caption{
 \label{fig:scatter_Qtf}
 \small{The relation Eq.~(\ref{eq:PQP-Q}) remains applicable for more general protocols than the TSP and for times $t_f$ comparable to the relaxation time of the system ($t_f\ll\tau_{relax}=1$).}}

%The relation predict by the natural logarithm of Eq.~(\ref{eq:PQP-Q}) continues to be valid even for $t_f\sim 1$ as is shown for the plot. The plot is based on $10^6$ simulations with parameters $t_f=0.7$, $k_i=1/2$, and $k_f=1$.}}
\end{figure}

By following a similar methodology as with the work distribution, we can obtain the characteristic function for the heat $\hat{P}_Q(z)=(1+b_1 z+b_2 z^2)^{-1/2}$  with the coefficients
\begin{eqnarray*}
  b_1&=&\frac{-i(k_m-k_i)k_m}{k_i}\sigma_m^{2}(t),\\
  b_2&=&\frac{ k_m^{2}}{k_i}\sigma_m^{2}(t),
\end{eqnarray*}
where $\sigma_m(t)$ is given by Eq.~(\ref{sigma m}). From this characteristic function, we can derive the average heat
\begin{equation}\label{eq:heat}
\expval{Q}_t=-\frac{k_m-k_i}{2 k_i}\left(1- e^{-2k_m t}\right),
\end{equation}
and its variance
\begin{equation}
\sigma_Q(t)^{2}=2\expval{Q}_t^2-\frac{2 k_m\expval{Q}_t}{k_m-k_i}.
\end{equation}
It is important to observe that the average heat and variance exhibit a characteristic time $\tau_m=1/k_m$. The characteristic function $\hat{P}_Q(z)$ has a similar structure to that of the work probability density function for $t > t_f$. Therefore, the inverse Fourier transform can be computed explicitly, leading to the heat probability distribution for $0\leq t \leq t_f$,
\begin{widetext}    
\begin{align}
    %P(Q,t)=&\frac{1}{\pi}\sqrt{\frac{k_i-k_m}{2k_m\expval{Q}_t}}\exp{\frac{k_i-k_m}{2k_m} Q}K_0\left(\frac{(k_i-k_m)\sqrt{\sigma_Q^2-\expval{Q}_t^2}}{2k_m\expval{Q}_t}\abs{Q}\right)\\
    \label{eq:PQ_TSP}
    P_{Q_D}(Q,t)&=\frac{1}{\pi\sqrt{\sigma_Q(t)^2-2\expval{Q}_t^2}}\exp{\frac{\expval{Q}_t}{\sigma_Q(t)^2-2\expval{Q}_t^2}Q}K_0\left(\frac{\sqrt{\sigma_Q(t)^2-\expval{Q}_t^2}}{\sigma_Q(t)^2 -2\expval{Q}_t^2}\abs{Q}\right).
\end{align}
\end{widetext}
This result can also be obtained using the methods described in Ref.~\cite{imparato}.
It is worth noting that the exponential term in the heat distribution is time-independent since $\frac{\expval{Q}_t}{\sigma_Q^2-2\expval{Q}_t^2}=\frac{k_i-k_m}{2k_m}$ does not depend on $t$. On the other hand, the argument of the Bessel function $K_0$ is time-dependent. The ratio of the heat probability distributions $P_{Q_D}(Q,t)/P_{Q_D}(-Q,t)$ satisfies a similar relation to the one found for the work in Eq.~(\ref{eq:PWP-W})
\begin{equation}\label{eq:PQP-Q}
    \frac{P_{Q_D}(Q,t)}{P_{Q_D}(-Q,t)}
    =\exp{\frac{2\expval{Q}_t}{{\sigma_Q^2}_t-2\expval{Q}_t^2}Q}=
    \exp{\frac{k_i-k_m}{k_m}Q}.
\end{equation}
A numerical test of this relation for non-TSP protocols in the range $t_f\ll \taurel=1$ is shown in Fig.~\ref{fig:hypothesis_Q} revealing again its universality. For longer protocol times $t_f \sim 1$, the relation given by Eq.~(\ref{eq:PQP-Q}), which has been proven for the TSP protocol, remains valid at large values of $|Q|$ but shows some deviations at small values as illustrated in Fig.~\ref{fig:scatter_Qtf} for other non-TSP protocols. 
The ratio $P_{Q_D}(Q,t)/P_{Q_D}(-Q,t)$ also appears in a different context when one considers the relaxation from the microcanonical to the canonical ensembles of this system~\cite{JDM23}.

%A similar relation to Eq.~(\ref{eq:PQP-Q}) where $P_{Q_D}(Q,t)/P_{Q_D}(-Q,t)$ is a simple exponential function of $Q$ also appears in a different context when one considers the relaxation from the microcanonical to the canonical ensembles of this system~\cite{JDM23}.

%{\em
%\gt{XXX: Por hacer: gráficas de esta relación comparando con los otros protocolos
%}
%}

\begin{figure}[t]
	\begin{center}
		\includegraphics[scale=0.6]{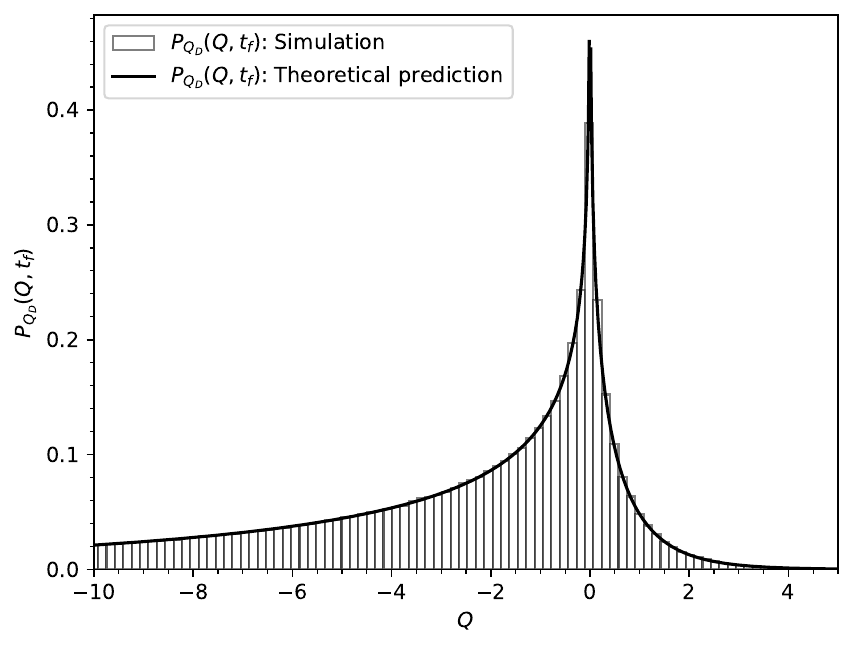}
	\end{center}
	\caption{
 \label{fig:heat_pdf}
 \small The probability density function for heat Eq.~(\ref{eq:PQ_TSP}) at $t = t_f$ and the numerical results for $10^6$ simulations. The simulations were conducted using a specific parameter setting, where $k_i=1/2$, $k_f=1$, and $t_f=1/30$.}
 %Notably, a remarkable level of agreement is observed between the theoretical and simulation outcomes, indicating a high degree of concurrence.}
\end{figure}

\subsubsection{For $t>t_f$}

After the second step of the TSP, the average heat reaches its equilibrium value given by Eq.~(\ref{eq:heat}) evaluated at $t=t_f$. However, surprisingly, moments of order $n\geq 2$ of the heat distribution take more time to reach their equilibrium values. To investigate this phenomenon, we attempt to calculate the heat distribution for $t>t_f$. This has the expression
\begin{equation}
P_{Q_D}(Q,t)=\int_{-\infty}^{\infty} dx_i dx_f dx_t\delta(Q-Q_D)P_3(x_i,t_i;x_f,t_f;x_t,t)
\end{equation}
which involves the JPD of the three-point $P_3(x_i,t_i;x_f,t_f;x_t,t)$. This quantity, is a product of two Ornstein-Uhlenbeck processes with stiffness $k_m$ and $k_f$ and the initial probability distribution, 
\begin{align}
    P_3(x_i,t_i;x_f,t_f;x_t,t) = &
    P(x_i,t_i)P(x_f,t_f|x_i,t_i)
    \nonumber\\
    & \times P(x_t,t|x_f,t_f),
\end{align}
with 
\begin{equation}
\label{eq:OH-kf}
    P(x_t,t|x_f,t_f)=\frac{1}{\sqrt{2\pi \sigma_f(t)^2}}\exp{-\frac{(x_t-x_fe^{-k_f(t-t_f)})^2}{2\sigma_f(t)^2}},
\end{equation}
where 
\begin{equation}
    \sigma_f(t)^2=\frac{1-e^{-2k_f (t-t_f)}}{k_f}.
\end{equation}
The transition probability $P(x_f,t_f|x_i,t_i)$ and the initial distribution are given by Eq.~(\ref{eq:OH-km}) and Eq.~(\ref{pdf initial}), respectively.

\begin{figure}[t]
	\begin{center}
		\includegraphics[scale=0.6]{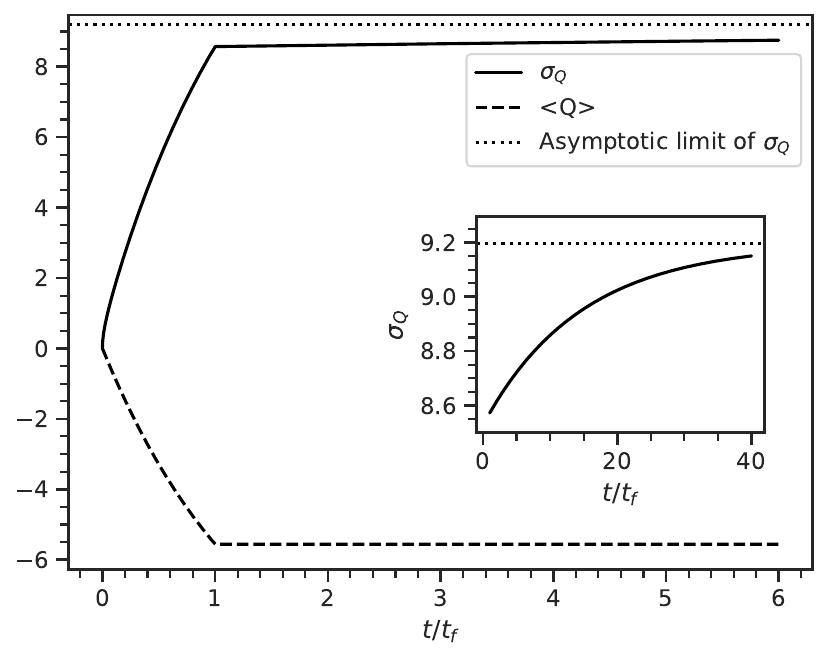}
	\end{center}
	\caption{
 \label{fig:heat_metrics}
 \small Standard deviation and average of heat for the TSP with parameters $t_f=1/30$, $k_i=1/2$ and $k_f=1$. The dotted line is the asymptotic value of $\sigma_Q$ when $t\rightarrow\infty$. The inset plot is a zoom of the variation of the function $\sigma_Q$ for $t>t_f$. } 
\end{figure}

By combining these expressions, utilizing the Fourier representation of the delta distribution, and performing the resulting Gaussian integrals, we can obtain the characteristic function
\begin{equation}\label{cubic}
    \hat{P}_{Q_D}(z)=\frac{1}{\sqrt{1+c_1 z+c_2 z^2+c_3 z^3}}
\end{equation}
whose coefficients are given by
\begin{eqnarray*}
  c_1&=&A(1-e^{-2k_f(t-t_f)})-\frac{ik_m(k_m-k_i)}{k_i}\sigma_m^2(t_f),\\
  c_2&=&B(1-e^{-2k_f(t-t_f)})+\frac{ k_m^2}{k_i}\sigma_m^{2}(t_f),\\
  c_3&=&C(1-e^{-2k_f(t-t_f)}),
\end{eqnarray*}
where $A, B$ and $C$ are constants given by
\begin{eqnarray*}
    A&=&\frac{i(k_i-k_f)}{k_i}+\frac{ik_f(k_m-k_i)}{k_i}\sigma_m^2(t_f),\\
    B&=&\frac{k_f+(k_fk_i-2k_mk_f-k_ik_m+k_m^2)\sigma_m^{2}(t_f)}{k_i},\\
    C&=&\frac{i(k_m-k_f)k_m\sigma_m^2(t_f)}{k_i}.
\end{eqnarray*}
These coefficients are completely general, meaning that $k_m$ does not necessarily satisfy the consistency equation Eq.~(\ref{consistency}). If $k_m$ satisfies the consistency equation, the coefficient $A$ vanishes but the coefficients $B$ and $C$ are nonzero. As a result, the average heat for $t > t_f$ becomes time-independent. However, the variance and higher-order moments of the heat PDF depend on time and have a characteristic relaxation time given by $1/k_f$. If $k_m$ is a solution of Eq.~(\ref{consistency}), the average heat and variance are given by
\begin{eqnarray}
  \expval{Q}&=&-\frac{k_m}{2}\left(\frac{1}{k_i}-\frac{1}{k_f}\right),\\
  \sigma_Q^{2}&=&2\expval{Q}^2-\frac{2k_m\expval{Q}}{k_m-k_i}+D(1-e^{-2k_f(t-t_f)}),
\end{eqnarray}
where $D$ is 
\begin{equation}
     D=\frac{(k_f-k_m)(k_i^2+k_fk_m-k_ik_m)}{k_ik_f(k_i-k_m)}.
\end{equation}
Figure~\ref{fig:heat_metrics} shows the evolution of the average heat and its standard deviation. When $t>t_f$, the average heat has stabilized, but its standard deviation continues to change in time with a relaxation time of order $1/k_f$. This finding is interesting, as it indicates that the position and work probability distribution have reached a state of equilibrium at the final time $t_f$. However, the heat distribution requires additional time to reach its ultimate state of equilibrium. The primary aim of the shortcut protocols is to achieve equilibrium in the position distribution, while the attainment of equilibrium in other relevant distributions is not necessarily simultaneous with that of the position distribution. 

\subsection{Produced entropy distribution for TSP}\label{sub_sec:entropy}
In this section, we specifically focus on the produced entropy associated with the two-step protocol. This quantity holds significant importance in understanding the inherent irreversibility within non-equilibrium systems and provides valuable insights into the fundamental thermodynamic principles that govern the TSP.

The Gibbs-Von Neumann entropy of the particle (with $k_B=1$) is defined as
\begin{equation}
    S(t) = -\int dx P(x,t)\ln{P(x,t)}
\end{equation}
where $P(x,t)$ is the solution of the Fokker-Planck equation for a given initial probability distribution $P_i(x)$. This entropy can be interpreted as the average of the quantity $-\ln{P(x,t)}$. This trajectory-dependent expression has information on the ensemble through the initial distribution \cite{seifert,jarzynski_equilibrium_1997}. Therefore, we can associate an entropy with a particular realization of the stochastic process $x(t)$ as
\begin{equation}
    s(t)=-\ln{P(x_t,t)}.
\end{equation}
From this, we can compute the particle entropy change (``the system" entropy change) from $t=0$ to $t$
\begin{equation}
    \Delta s =s(t)-s(0)=\ln{\frac{P_i(x_i)}{P(x_t,t)}}.
\end{equation}
Hence, the total entropy change of the system $+$ environment is
\begin{equation}\label{deltaS}
\Delta S = -\frac{Q_D}{T} +\ln{\frac{P_i(x_i)}{P(x_t,t)}}.
\end{equation}
%where $\Delta S$ is the entropy change between two states with the probability density function (PDF) of the position given by $P_i$ and $P_f$, respectively. 
The first term in Eq.~(\ref{deltaS}) corresponds to the increase in entropy of the environment, which is assumed to be in equilibrium at the temperature $T$ (in dedimensionalized units $T=1$) and the last term represents the increase in entropy of the system. We will use $\Delta S$ and $\Sigma_D$ interchangeably to represent produced entropy.

By employing a methodology analogous to that used to calculate the probability density functions of heat and work, we can derive the PDF of produced entropy. Due to the similarities with the previous derivations, we will omit several intermediate steps.
Once again, it is necessary to consider two distinct cases.

\subsubsection{For $0< t< t_f$}
To calculate the system entropy change, we need the probability density function of the position at the initial time as well as the position PDF at $t$, as given by Eq.~(\ref{position pdf}). By utilizing the expression for the heat transferred from the environment to the particle in the process, as shown in Eq.~(\ref{eq:heat done}), and the two-point probability distribution given in Eq.~(\ref{two points pdf}), we can determine the total entropy change (with $T=1$) for $t \leq t_f$ as follows
\begin{equation}
    \Sigma_D =\frac{1}{2}k_m(x_i^2-x_t^2)+\Sigma_0(t)-\frac{x_i^2}{2\sigma_i^2}+\frac{x_t^2}{2\sigma_X(t)^2},
\end{equation}
where $ \Sigma_0(t)=\frac{1}{2}\ln{\frac{\sigma_X(t)^2}{\sigma_i^2}}$ is a deterministic function of time.

Similar to the computation of work and heat distributions, the PDF of produced entropy can be expressed as:
\begin{equation}\label{eq:entropy_repre}
P_{\Sigma_D}(\Sigma) = \frac{1}{2\pi}\int_{-\infty}^{\infty} dz \hat{P}_{\Sigma_D}(z) \exp[iz(\Sigma-\Sigma_0)],
\end{equation}
where $\hat{P}_{\Sigma_D}(z)e^{-iz\Sigma_0}$ is the characteristic function of the entropy and the function $\hat{P}_{\Sigma_D}(z)$ is given by
\begin{equation}
    \hat{P}_{\Sigma_D}(z)=\frac{1}{\sqrt{1+d_1z+d_2z^2}},
\end{equation}
whose coefficients are
\begin{eqnarray}
  d_1&=&\frac{i \left(-k_m \sigma_m ^4 k_i+k_m^2 \sigma_m ^4-k_m \sigma_m ^2+1\right)}{\sigma_m ^2 k_i},\\
  d_2&=&\frac{\left(k_m-k_i\right) \left(k_m \sigma_m ^2-1\right)}{k_i}.
\end{eqnarray}
Computing the integral Eq.~(\ref{eq:entropy_repre}), we arrive at the expression
\begin{widetext}    
\begin{align}\label{pdf entropy}
    %P(Q,t)=&\frac{1}{\pi}\sqrt{\frac{k_i-k_m}{2k_m\expval{Q}_t}}\exp{\frac{k_i-k_m}{2k_m} Q}K_0\left(\frac{(k_i-k_m)\sqrt{\sigma_Q^2-\expval{Q}_t^2}}{2k_m\expval{Q}_t}\abs{Q}\right)\\
    P_{\Sigma_D}(\Sigma,t)&=\frac{1}{\pi\sqrt{\sigma_{ \Sigma}(t)^2-2\expval{ \Sigma- \Sigma_0}_t^2}}\exp{\frac{\expval{ \Sigma-\Sigma_0}_t}{\sigma_{ \Sigma}(t)^2-2\expval{ \Sigma - \Sigma_0}_t^2}(\Sigma - \Sigma_0)} K_0\left(\frac{\sqrt{\sigma_{ \Sigma}(t)^2-\expval{  \Sigma - \Sigma_0}_t^2}}{\sigma_{ \Sigma}(t)^2 -2\expval{ \Sigma - \Sigma_0}_t^2}\abs{  \Sigma - \Sigma_0}\right),
\end{align}
\end{widetext}
where 
\begin{eqnarray}
  \expval{\Sigma}_t&=&\Sigma_0(t)+\frac{k_m-k_i}{2k_i}\left(1-e^{-2k_mt}\right),
  \label{eq:sigmat}
  \\
  \sigma_\Sigma^{2}(t)&=&2\expval{ \Sigma - \Sigma_0}_t^2+\frac{(k_i-k_m)^2\left(1-e^{-2k_mt}\right)}{k_i\left(k_m-k_i\left(1-e^{2k_mt}\right)\right)}.
  \label{eq:sigmavart}
\end{eqnarray}

\subsubsection{For $t> t_f$}

In this case, the entropy is given by
\begin{equation}
\Sigma_D = \frac{1}{2}k_m(x_i^2-x_f^2) + \frac{1}{2}k_f(x_f^2-x_t^2) + \Sigma_0 - \frac{x_i^2}{2\sigma_i^2} + \frac{x_t^2}{2\sigma_X(t)^2}.
\end{equation}
It is worth noting that if we ensure a final equilibrium situation at $t_f$ by imposing the consistency relation Eq.~(\ref{consistency}), there is a cancellation between $-k_f x_t^2/2$ and $x_t^2/(2\sigma_X(t)^2)$ making $\Sigma_D$ time-independent.

Assuming that $k_m$ is  arbitrary, the function $\hat{P}_{\Sigma_D}(z)$ takes the form
\begin{equation}\label{eq:f-coe}
\hat{P}_{\Sigma_D}(z) = \frac{1}{\sqrt{1 + f_1z + f_2z^2 + f_3z^3}},
\end{equation}
where the coefficients $f_1, f_2,$ and $f_3$ are complex expressions of the system parameters (refer to Appendix~\ref{app:2} for details). Due to the cubic polynomial inside the square root, the integral in Eq.~(\ref{eq:entropy_repre}) cannot be evaluated in closed form.

\begin{figure}[t]
	\begin{center}
		\includegraphics[scale=0.6]{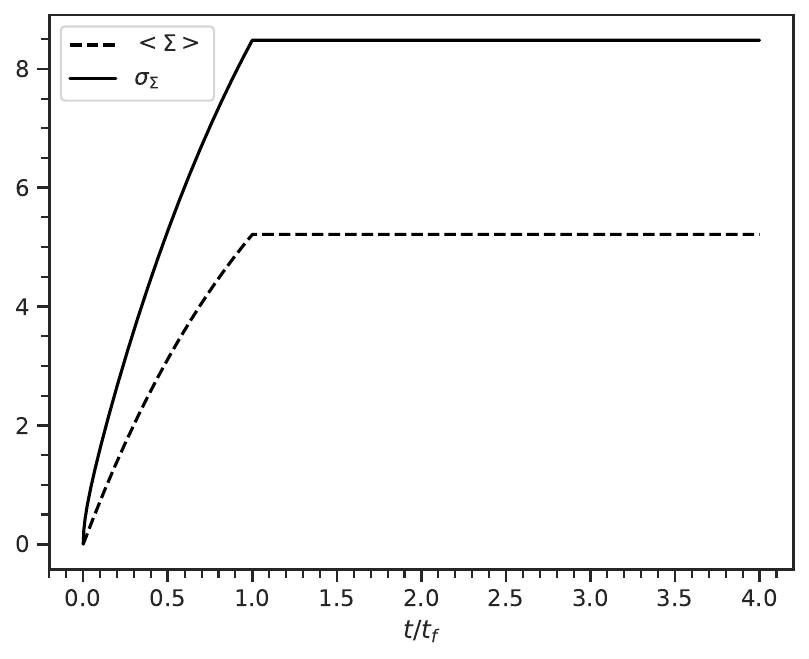}
	\end{center}
	\caption{
 \label{fig:entropy_metrics}
 \small Standard deviation and average for the produced entropy for the TSP protocol with parameters $k_i=1/2$, $k_f=1$ and $t_f=1/30$.}
\end{figure}

Therefore, in this section we will impose the constraint that $k_m$ satisfies the consistency equation Eq.~(\ref{consistency}). Under this constraint, the coefficient $f_3$ vanishes, allowing us to obtain a closed-form expression. The result is similar to the result in the previous time interval Eq.~(\ref{pdf entropy}) although the average and variance will exhibit modifications. More specifically, they are time-independent (as expected for an equilibrium situation) and are given by
\begin{eqnarray}
  \expval{\Sigma}&=&\Sigma_{0,f}+\frac{k_m}{2}\left(\frac{1}{k_i}-\frac{1}{k_f}\right),\\
  \sigma_\Sigma^{2}&=&2\expval{\Sigma-\Sigma_{0,f}}^2+\frac{2\expval{\Sigma-\Sigma_{0,f}}(k_m-k_f)}{k_m},
\end{eqnarray}
with $\Sigma_{0,f}=\Sigma_0(t_f)=\frac{1}{2}\ln(k_i/k_f)$. This average and variance of $\Sigma_D$ coincide with the final values obtained in the previous time interval, Eqs.~(\ref{eq:sigmat}) and~(\ref{eq:sigmavart}) at $t=t_f$. 
%Furthermore, for $t > t_f$, these values remain constant as expected for an equilibrium situation. 
Figure~\ref{fig:entropy_metrics} shows the evolution of the average produced entropy and its standard deviation.

\begin{figure}[t]
	\begin{center}
		\includegraphics[scale=0.6]{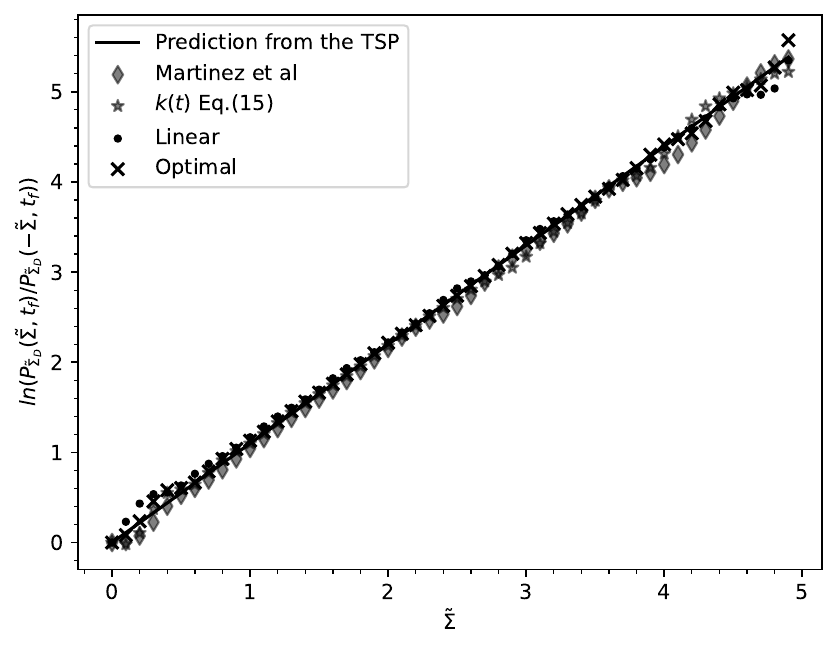}
	\end{center}
	\caption{
 \label{fig:PSP-S}
 \small Plot of the Eq.~(\ref{eq:psp-s}) for the TSP protocol and simulation results for the other protocols. $10^6$ simulations have been performed with $k_i=1/2$, $k_f=1$ and $t_f=1/30$.}
\end{figure}

\begin{figure}[t]
	\begin{center}
		\includegraphics[scale=0.6]{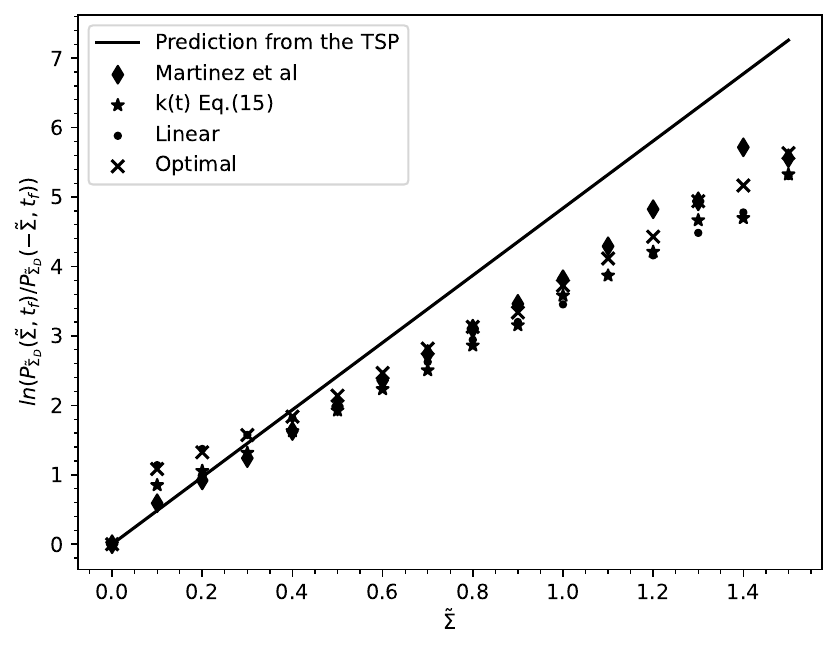}
	\end{center}
	\caption{
 \label{fig:scattertf}
 \small The prediction given by Eq.~(\ref{eq:psp-s}) fails when $t_f\sim 1$; however, the behavior looks linear. This plot is for $10^6$ simulations with parameters $k_i=1/2$, $k_f=1$ and $t_f=0.7$. }
\end{figure}

The produced entropy distribution satisfies a relation similar to the ones for work and heat shown in Eqs.~(\ref{eq:PWP-W}) and~(\ref{eq:PQP-Q}) if we define $\tilde\Sigma=\Sigma-\Sigma_0$. This is given by
\begin{eqnarray}
     \frac{P_{\tilde\Sigma_D}(\tilde\Sigma)}{P_{\tilde\Sigma_D}(-\tilde\Sigma)}&=&\exp{\frac{2\expval{\Sigma-\Sigma_0}_t}{\sigma_\Sigma^2-2\expval{\Sigma-\Sigma_0}_t^2}\tilde\Sigma}
     \nonumber\\
     &=&\exp{\frac{k_m}{k_m-k_f}\tilde\Sigma}
    \,.\label{eq:psp-s}
\end{eqnarray} 
Figure~\ref{fig:PSP-S} represents this relation for $t=t_f\ll 1$ and the simulation results for the non-TSP protocols considered in this paper. This plot shows the universality of the relation Eq.~(\ref{eq:psp-s}) when $t_f\ll 1$. However, this universality breaks down for longer protocol times $t_f\sim 1$ as can be observed in Fig.~\ref{fig:scattertf}.

\subsubsection{Fluctuation theorem for the produced entropy}
The entropy distribution must satisfy a fluctuation theorem ~\cite{Evans1, Evans2}, in the same way, that the work distribution must satisfy the Jarzynski equality and the Crocks relation. Hence, we will prove the integral fluctuation relation for the produced entropy.
Using the definition of the expectation value and inserting the integral representation given in Eq.~(\ref{eq:entropy_repre}), we can obtain the following identity 
\begin{equation}
\langle e^{-\Sigma} \rangle = \hat{P}_{\Sigma_D}(-i)e^{-\Sigma_0(t)}.
\end{equation}
%with $ \Sigma_0=\frac{1}{2}\ln{\frac{\sigma_X(t)^2}{\sigma_i^2}}$. 
For arbitrary $k_m$ (not necessarily a solution of Eq.~(\ref{consistency})), we can compute $\hat{P}_{\Sigma_D}(-i)$ as
\begin{equation}
\hat{P}_{\Sigma_D}(-i) = \frac{\sigma_X(t)}{\sigma_i}.
\end{equation}
As a result, the expectation value is given by
\begin{equation}
\langle e^{-\Sigma} \rangle = 1.
\end{equation}
This is the well-known integral fluctuation theorem for produced entropy~\cite{seifert,Evans2,gallavottiDynamicalEnsemblesStationary1995}.

\section{Conclusion}

We have reviewed how to construct ESE protocols. In particular, we have shown that a Brownian particle in a harmonic potential with time-dependent stiffness and at initial equilibrium with a Gaussian distribution, the position distribution remains Gaussian for later times. Additionally, the variance of the position distribution depends on the protocol. 
Moreover, we have thoroughly described the two-step protocol and examined its energetic behaviors. Notably, we obtained analytical expressions for the distribution functions of work, heat, and produced entropy. These analytical solutions comply with the fluctuation theorems and identities well-established in the existing literature Refs.~\cite{Evans1, Evans2, gallavottiDynamicalEnsemblesStationary1995, Crooks1998, crooks_entropy_1999, Jar97, jarzynski_equilibrium_1997}.

The significance of the two-step protocol extends beyond analytical solutions, as it also characterizes more general protocols within the time range of interest of the shortcut protocols (i.e., when $t_f \ll \taurel$). Our findings unveil intriguing insights into the behavior of various protocols that are of common interest for theoretical and experimental research. The average value of the stiffness $k_m$ is inversely proportional to the protocol duration $t_f$, explicitly given by Eq.~(\ref{eq:km_tf_short}). This relation can serve as a guide for designing fast thermalization protocols. It also imposes some restrictions on them. If for some practical application the stiffness has to be bounded, Eq.~(\ref{eq:km_tf_short}) imposes a limit on how short the protocol can be~\cite{trizac3}. Regarding the energetics statistics, we obtained strong numerical evidence that suggests that the probability distribution functions of work, heat, and produced entropy satisfy universal simple relations given by Eqs.~(\ref{eq:PWP-W}), ~(\ref{eq:PQP-Q}) and~(\ref{eq:psp-s}) when we compare the positive versus the negative production of these quantities.

Despite the valuable insights gained, this study faced certain limitations, particularly in the context of the overdamped regime. Future research may involve finding analytical solutions in the underdamped regime and extending the analysis to systems of interacting particles. By addressing these aspects, future studies can provide a more comprehensive understanding of the fast thermalization protocols for Brownian particles in complex environments, further enriching the field of stochastic thermodynamics.

In summary, our research significantly advances the comprehension of stochastic thermodynamics in colloidal Brownian particles. We have gained valuable insights into the dynamics of overdamped Brownian particles under time-dependent potentials and external control, which facilitates the acceleration of equilibration times. These findings hold great potential for the design and application of efficient protocols across diverse scientific and engineering disciplines.

\section{Acknowledgments}
This work was supported by Fondo de Investigaciones, Facultad de Ciencias,
Universidad de los Andes INV-2021-128-2267. We thank Emmanuel Trizac for the fruitful discussions.

\section{Appendixes}
\appendix 
\section{}\label{app:1}
In order to prove the Crooks identity Eq.~(\ref{eq:crooks identity}), it is necessary to demonstrate the symmetry under the permutation $k_i \leftrightarrow k_f$ of the Bessel function in Eq.~(\ref{eq:pdf work}). To achieve this goal, we use the following definitions,
\begin{eqnarray}
 S &=& \frac{(k_m-k_f)(k_m-k_i)\sigma_m^2(t)}{2},\\
 A &=& \frac{k_f-k_i}{2},
\end{eqnarray}
where $S$ and $A$ stand for symmetric and antisymmetric under the permutation, respectively.
Eqs.~(\ref{metrics general}) and~(\ref{eq:metrics-gen-sigma}) take the form
\begin{eqnarray}
 \expval{W} &=& \frac{A+S}{k_i},\\
 \sigma_W^2 &=& \frac{2(A+S)^2+2k_iS}{k_i^2}.
\end{eqnarray}
Now, the combination in the argument of the Bessel function is 
\begin{equation}
    \frac{\sqrt{\sigma_W^2-\expval{W}^2}}{\sigma_W^2-2\expval{W}^2}=\frac{\sqrt{A^2+S^2+2S(A+k_i)}}{2S}.
\end{equation}
The only term that is not evident symmetric is the combination $A+k_i$, however
\begin{equation}
    A+k_i=\frac{k_i+k_f}{2}
\end{equation}
that is symmetric under the permutation.

\section{}\label{app:2}
In this appendix, we give the intricate expressions of the coefficients given intervening in Eq.~(\ref{eq:f-coe}):
\begin{eqnarray*}
    \sigma _t^2f_3 &=& i \sigma _f^2 \sigma _m^2 \left(k_f-k_m\right) \left(k_f \sigma _t^2-1\right) \left(\sigma _i^2 k_m-1\right),\\
    \sigma _t^2f_2 &=& \sigma _m^2 \left(\sigma _i^2 k_m-1\right) \left(k_m \sigma _t^2-1\right)
    -\sigma _f^2 \left(k_f \sigma _t^2-1\right)\\
    &&\times\left(\sigma _m^2 \left(2 k_f-k_m\right) \left(\sigma _i^2 k_m-1\right)-k_f \sigma _i^2+1\right),\\
    \sigma _t^2f_1 &=& -i\sigma _f^2 \left(k_f \sigma _t^2-1\right) \left(k_f \left(\sigma _m^2 \left(\sigma _i^2 k_m-1\right)-\sigma _i^2\right)+1\right)+\\
    &+&i\sigma _i^2 \left(k_m \sigma _m^2 \left(k_m \sigma _t^2-1\right)+1\right)-i\sigma _t^2 \left(k_m \sigma _m^2+1\right)+\\
    &+&i\sigma _m^2,
\end{eqnarray*}
where $\sigma_t^2$ is the solution of Eq.~(\ref{n2}) for the variance $\chi_2(t)$ in the interval $t\geq t_f$ given by
\begin{equation}
    \sigma^2_t=\left(-\frac{1}{k_f}+\frac{1-k_m \sigma _m^2}{k_i}+\sigma _m^2\right)e^{-2 k_f \left(t-t_f\right)}+\frac{1}{k_f}.
\end{equation}
Substituting the consistency equation Eq.~(\ref{consistency}), these coefficients collapse to
\begin{eqnarray}
    f_3 &=&0, \\
    f_2 &=& \frac{\left(k_i-k_f\right) \left(k_f-k_m\right)}{k_f k_i},\\
    f_1 &=& -i k_m \left(\frac{1}{k_f}-\frac{1}{k_i}\right).
\end{eqnarray}
Because $f_3$ is zero, the polynomial Eq.~(\ref{eq:f-coe}) is of second order and the inverse Fourier transform can be computed in terms of Bessel functions, yielding Eq.~(\ref{pdf entropy}).

\nocite{*}

\bibliography{TSP_article} % Produces the bibliography via BibTeX.

\end{document}